\documentclass[twocolumn,tight,dvipsnames,twocolappendix]{aastex63}

\usepackage{amsmath}

\received{XX}
\revised{YY}
\accepted{ZZ}
\submitjournal{ApJ}

\shorttitle{MAPS II: Imaging Strategies}
\shortauthors{Czekala et al.}

\newcommand{\clean}{CLEAN}
\newcommand{\tclean}{\texttt{tclean}}
\newcommand{\rml}{RML}

\newcommand\kms{\ifmmode{\rm km\thinspace s^{-1}}\else km\thinspace s$^{-1}$\fi}

\newcommand{\jypixel}{$\mathrm{Jy}\thinspace\{\mathrm{pixel}\}^{-1}$}
\newcommand{\jybeam}{$\mathrm{Jy}\thinspace\{\mathrm{beam}\}^{-1}$}
\newcommand{\jycleanbeam}{$\mathrm{Jy}\thinspace\{\mathrm{\clean\ beam}\}^{-1}$}
\newcommand{\jydirtybeam}{$\mathrm{Jy}\thinspace\{\mathrm{dirty\ beam}\}^{-1}$}
\newcommand{\jyarcsec}{$\mathrm{Jy}\thinspace\{\mathrm{arcsec}\}^{-2}$}

\newcommand{\uv}{$uv$}

\usepackage[version=3]{mhchem}

\begin{document}

\title{Molecules with ALMA at Planet-forming Scales (MAPS) II: CLEAN Strategies for Synthesizing Images of Molecular Line Emission in Protoplanetary Disks}

\correspondingauthor{Ian Czekala}
\email{iczekala@psu.edu}

\author[0000-0002-1483-8811]{Ian Czekala}
\altaffiliation{NASA Hubble Fellowship Program Sagan Fellow}
\affiliation{Department of Astronomy and Astrophysics, 525 Davey Laboratory, The Pennsylvania State University, University Park, PA 16802, USA}
\affiliation{Center for Exoplanets and Habitable Worlds, 525 Davey Laboratory, The Pennsylvania State University, University Park, PA 16802, USA}
\affiliation{Center for Astrostatistics, 525 Davey Laboratory, The Pennsylvania State University, University Park, PA 16802, USA}
\affiliation{Institute for Computational \& Data Sciences, The Pennsylvania State University, University Park, PA 16802, USA}
\affiliation{Department of Astronomy, 501 Campbell Hall, University of California, Berkeley, CA 94720-3411, USA}

\author[0000-0002-8932-1219]{Ryan A. Loomis}
\affiliation{National Radio Astronomy Observatory, 520 Edgemont Rd., Charlottesville, VA 22903, USA}

\author[0000-0003-1534-5186]{Richard Teague}
\affiliation{Center for Astrophysics \textbar\, Harvard \& Smithsonian, 60 Garden St., Cambridge, MA 02138, USA}

\author[0000-0003-2014-2121]{Alice S. Booth}
\affiliation{Leiden Observatory, Leiden University, 2300 RA Leiden, the Netherlands}
\affiliation{School of Physics and Astronomy, University of Leeds, Leeds, LS2 9JT, UK}

\author[0000-0001-6947-6072]{Jane Huang}
\altaffiliation{NHFP Sagan Fellow}
\affiliation{Department of Astronomy, University of Michigan, 323 West Hall, 1085 S. University Avenue, Ann Arbor, MI 48109, USA}
\affiliation{Center for Astrophysics \textbar\, Harvard \& Smithsonian, 60 Garden St., Cambridge, MA 02138, USA}

\author[0000-0002-2700-9676]{Gianni Cataldi}
\affil{National Astronomical Observatory of Japan, Osawa 2-21-1, Mitaka, Tokyo 181-8588, Japan}
\affil{Department of Astronomy, Graduate School of Science, The University of Tokyo, Tokyo 113-0033, Japan}

\author[0000-0003-1008-1142]{John~D.~Ilee}
\affil{School of Physics \& Astronomy, University of Leeds, Leeds LS2 9JT, UK}

\author[0000-0003-1413-1776]{Charles J. Law}
\affiliation{Center for Astrophysics \textbar\, Harvard \& Smithsonian, 60 Garden St., Cambridge, MA 02138, USA}

\author[0000-0001-6078-786X]{Catherine Walsh}
\affiliation{School of Physics and Astronomy, University of Leeds, Leeds, LS2 9JT, UK}

\author[0000-0003-4001-3589]{Arthur D. Bosman}
\affiliation{Department of Astronomy, University of Michigan, 323 West Hall, 1085 S. University Avenue, Ann Arbor, MI 48109, USA}

\author[0000-0003-4784-3040]{Viviana V. Guzm\'{a}n}
\affiliation{Instituto de Astrof\'isica, Pontificia Universidad Cat\'olica de Chile, Av. Vicu\~na Mackenna 4860, 7820436 Macul, Santiago, Chile} 

\author[0000-0003-1837-3772]{Romane Le Gal}
\affiliation{Center for Astrophysics \textbar\, Harvard \& Smithsonian, 60 Garden St., Cambridge, MA 02138, USA}
\affiliation{IRAP, Universit\'{e} de Toulouse, CNRS, CNES, UT3, 31400 Toulouse, France}
\affiliation{IPAG, Universit\'{e} Grenoble Alpes, CNRS, IPAG, 38000 Grenoble, France}
\affiliation{IRAM, 300 rue de la piscine, F-38406 Saint-Martin d'H\`{e}res, France}

\author[0000-0001-8798-1347]{Karin I. \"Oberg}
\affiliation{Center for Astrophysics \textbar\, Harvard \& Smithsonian, 60 Garden St., Cambridge, MA 02138, USA}

\author[0000-0003-4099-6941]{Yoshihide Yamato}
\affiliation{Department of Astronomy, Graduate School of Science, The University of Tokyo, Tokyo 113-0033, Japan}

\author[0000-0003-3283-6884]{Yuri Aikawa}
\affiliation{Department of Astronomy, Graduate School of Science, The University of Tokyo, Tokyo 113-0033, Japan}

\author[0000-0003-2253-2270]{Sean M. Andrews}
\affiliation{Center for Astrophysics \textbar\, Harvard \& Smithsonian, 60 Garden St., Cambridge, MA 02138, USA}

\author[0000-0001-7258-770X]{Jaehan Bae}
\altaffiliation{NASA Hubble Fellowship Program Sagan Fellow}
\affil{Earth and Planets Laboratory, Carnegie Institution for Science, 5241 Broad Branch Road NW, Washington, DC 20015, USA}
\affiliation{Department of Astronomy, University of Florida, Gainesville, FL 32611, USA}

\author[0000-0003-4179-6394]{Edwin A.\ Bergin}
\affiliation{Department of Astronomy, University of Michigan, 323 West Hall, 1085 S. University Avenue, Ann Arbor, MI 48109, USA}

\author[0000-0002-8716-0482]{Jennifer B. Bergner} 
\altaffiliation{NASA Hubble Fellowship Program Sagan Fellow}
\affiliation{University of Chicago Department of the Geophysical Sciences, Chicago, IL 60637, USA}

\author[0000-0003-2076-8001]{L. Ilsedore Cleeves}
\affiliation{Department of Astronomy, University of Virginia, Charlottesville, VA 22904, USA}

\author[0000-0002-2358-4796]{Nicolas T. Kurtovic}
\affiliation{Departamento de Astronom\'ia, Universidad de Chile, Camino El Observatorio 1515, Las Condes, Santiago, Chile}
\affiliation{Max-Planck-Institut f\"{u}r Astronomie, K\"{o}nigstuhl 17, 69117, Heidelberg, Germany}

\author[0000-0002-1637-7393]{Fran\c cois M\'enard}
\affiliation{IPAG, Universit\'{e} Grenoble Alpes, CNRS, IPAG, 38000 Grenoble, France}

\author[0000-0002-7058-7682]{Hideko Nomura}
\affiliation{National Astronomical Observatory of Japan, 2-21-1 Osawa, Mitaka, Tokyo 181-8588, Japan}

\author[0000-0002-1199-9564]{Laura M. Pérez}
\affiliation{Departamento de Astronomía, Universidad de Chile, Camino El Observatorio 1515, Las Condes, Santiago, Chile}

\author[0000-0001-8642-1786]{Chunhua Qi}
\affiliation{Center for Astrophysics \textbar\, Harvard \& Smithsonian, 60 Garden St., Cambridge, MA 02138, USA}

\author[0000-0002-6429-9457]{Kamber R. Schwarz}
\altaffiliation{NASA Hubble Fellowship Program Sagan Fellow}
\affiliation{Lunar and Planetary Laboratory, University of Arizona, 1629 E. University Blvd, Tucson, AZ 85721, USA}

\author[0000-0002-6034-2892]{Takashi Tsukagoshi} \affiliation{National Astronomical Observatory of Japan, 2-21-1 Osawa, Mitaka, Tokyo 181-8588, Japan}

\author[0000-0002-1566-389X]{Abygail R. Waggoner}
\affiliation{Department of Chemistry, University of Virginia, Charlottesville, VA 22904}

\author[0000-0003-1526-7587]{David J. Wilner}
\affiliation{Center for Astrophysics \textbar\, Harvard \& Smithsonian, 60 Garden St., Cambridge, MA 02138, USA}

\author[0000-0002-0661-7517]{Ke Zhang}
\altaffiliation{NASA Hubble Fellow}
\affiliation{Department of Astronomy, University of Wisconsin-Madison, 475 N Charter St, Madison, WI 53706}
\affiliation{Department of Astronomy, University of Michigan, 323 West Hall, 1085 S. University Avenue, Ann Arbor, MI 48109, USA}

\begin{abstract}
The Molecules with ALMA at Planet-forming Scales large program (MAPS LP) surveyed the chemical structures of five protoplanetary disks across more than 40 different spectral lines at high angular resolution (0\farcs15 and 0\farcs30 beams for Bands 6 and 3, respectively) and sensitivity (spanning 0.3 - 1.3\,m\jybeam\ and 0.4 - 1.9\,m\jybeam\ for Bands 6 and 3, respectively). In this article, we describe our multi-stage workflow---built around the CASA \tclean\ image deconvolution procedure---that we used to generate the core data product of the MAPS LP: the position-position-velocity image cubes for each spectral line. Owing to the expansive nature of the survey, we encountered a range of imaging challenges; some are familiar to the sub-mm protoplanetary disk community, like the benefits of using an accurate CLEAN mask, and others less well-known, like the incorrect default flux scaling of the CLEAN residual map first described in \citet{jorsater95} (the ``JvM effect'').  We distill lessons learned into recommended workflows for synthesizing image cubes of molecular emission. In particular, we describe how to produce image cubes with accurate fluxes via the ``JvM correction,'' a procedure that is generally applicable to any image synthesized via CLEAN deconvolution but is especially critical for low S/N emission. We further explain how we used visibility tapering to promote a common, fiducial beam size and contextualize the interpretation of signal to noise ratio when detecting molecular emission from protoplanetary disks. This paper is part of the MAPS special issue of the Astrophysical Journal Supplement.
\end{abstract}

\keywords{protoplanetary disks --- 
submillimeter astronomy --- radio interferometry --- deconvolution --- surveys}

\section{Introduction} \label{sec:intro}
Sub-mm interferometers like the Atacama Large Millimeter/submillimeter Array (ALMA) enable high spatial and spectral resolution observations of protoplanetary disks. The Molecules with ALMA at Planet-forming Scales large program (MAPS LP) used ALMA to survey the chemical structures of five protoplanetary disks across more than 40 different spectral lines: an overview of the program and references to the full suite of MAPS papers is provided in \citet{oberg20}. In this paper, we describe the imaging strategies we employed to synthesize position-position-velocity image cubes from the interferometric visibilities. These image cubes form a core data product of the MAPS LP from which other value-added data products like moment maps, radial intensity profiles, and emission surfaces are derived \citep{law20_rad,law20_surf}. 

Consider an astronomical source whose spatial sky brightness distribution is described by $I$. Over small angular extents, the distribution is indexed $I(l,m)$ by the direction cosines $l=\sin \left ( \Delta \alpha \cos \delta \right )$ and $m = \sin \left( \Delta \delta \right )$ relative to some phase center, where $\alpha$ is right ascension and $\delta$ is declination; $l$ increases to the east and $m$ increases to the north. The visibility function of an astronomical source is the Fourier transform of the sky brightness distribution, given by
\begin{equation}
    {\cal V}(u,v) = \iint I(l,m) \exp \left \{- 2 \pi i (ul + vm) \right \} \, \mathrm{d}l\,\mathrm{d}m,
    \label{eqn:fourier}
\end{equation}
and is a complex quantity having real and imaginary components with units of flux, e.g., Jy \citep[for a full discussion of the conditions that must be met in order for Equation~\ref{eqn:fourier} to be accurate, see][Ch.\ 3]{thompson17}. Interferometers sample ${\cal V}$ at a discrete set of spatial frequencies fundamentally dictated by the array configuration and observing frequency \citep{thompson17}. The spatial frequencies $(u,v)$ are measured in multiples\footnote{Though $l$ and $m$ are technically unitless, for small angular extent, they could also be considered to have units of radians, implying that $u$ and $v$ can also be interpreted in units of cycles per radian.} of the observing wavelength $\lambda$ or $\mathrm{k}\lambda$, which can also be converted to length (e.g., m), directly corresponding to the instantaneous, projected baselines of the array.\footnote{Following the convention for $l$ and $m$, $u$ increases to the east and $v$ increases to the north. Since $u$ and $v$ correspond to the baseline locations of the array, we plot $u$ increasing towards the right, as if the array were viewed on the surface of the earth from above \citep[e.g.,][Figure 3.2]{thompson17}.} For radio interferometers like ALMA, the fundamental data product is then---for every spectral channel---a set of calibrated visibility measurements at various $(u_k,v_k)$ coordinates. These visibilities are complex-valued numbers measured in the presence of Gaussian noise 
\begin{equation}
V_{\mathrm{data},k} = {\cal V}(u_k, v_k) + \varepsilon_k.
\end{equation}
The distribution from which the noise $\varepsilon_k$ is drawn is usually well-described by a Gaussian distribution with the thermal weights\footnote{Often these weights are presented proportionally $w_k \propto \sigma_k^{-2}$ such that the constant of proportionality is determined via the RMS noise on a single correlator channel corresponding to a visibility of unit weight \citep[\S3.2;][]{briggs95}. In our definition, we assume that this calibration has already been carried out and that the weights have a direct statistical interpretation.} equal to the inverse variance $w_k = \sigma_k^{-2}$. Observational data is often interpreted in the context of a model. Fortunately, this measurement process defines a straightforward likelihood function for any set of model visibilities $\boldsymbol{V}_\mathrm{model}$,
\begin{equation}
    \ln {\cal L} = \ln p(\boldsymbol{V}_\mathrm{data} | \boldsymbol{V}_\mathrm{model}) \propto -\frac{\chi^2}{2},
    \label{eqn:log-likelihood}
\end{equation}
where
\begin{equation}
    \chi^2 = \sum_k^N w_k \left | V_{\mathrm{data},k} - V_{\mathrm{model},k} \right |^2
\end{equation}
and $N$ is the number of visibility measurements. Throughout this work, we use bold notation to signify vector quantities. Model visibilities can be generated from analytic forward models $I(l,m) \leftrightharpoons {\cal V}(u,v)$ \citep[e.g., an axisymmetric intensity profile describing the thermal emission from dust rings in a protoplanetary disk;][]{zhang16,guzman18,jennings20} or by Fast-Fourier-transforming complete radiative transfer models of $I(l,m)$ (e.g., complicated dust morphologies (\citealt{tazzari18}) or kinematic models of spatially resolved CO emission (\citealt{czekala19})). Regardless of how model visibilities are generated, model fitting with the visibility likelihood function (Equation~\ref{eqn:log-likelihood}) and judicious choices of prior probability distributions creates a well-motivated posterior probability distribution that can be used for Bayesian parameter inference \citep[e.g.,][]{hogg18,speagle19}. 

Most of the difficulty in synthesizing \emph{images} representative of the sky brightness distribution $I(l,m)$ stems from the fact that the visibility function is not adequately sampled at all of the spatial frequencies where it has significant power. Because the Fourier transform is a linear operator, one maximum likelihood image solution is simply the inverse Fourier transform of the visibility measurements. In this inversion process, the unsampled spatial frequencies are typically set to zero power. The image that results is called the ``dirty image.''\footnote{Both because of its typically lousy aesthetics and because it forms the starting point for the \clean\ family of deconvolution algorithms.} By definition, the point spread function (PSF) for the dirty image is the system response to an impulse sky-brightness distribution ($I(l,m) = \delta(0,0)$, where in this context $\delta$ represents the Dirac delta function). The point spread function is also called the ``dirty beam'' because it typically has a substantial sidelobe pattern, responsible for the low fidelity of the dirty image. The PSF width and sidelobe pattern amplitude can be altered by adjusting the scheme by which nearby \uv\ points are averaged or ``gridded'' \citep[with tradeoffs against the thermal noise level;][]{briggs95}.

The PSF sidelobe response can be (effectively, albeit not perfectly) removed from the dirty image through image deconvolution. The most widely used deconvolution algorithm in the radio astronomy community is \clean\ \citep{hogbom74}, which we describe in \S\ref{sec:cleaning-process}. For more background on the \clean\ family of algorithms, see \citet[Ch.\ 11,][]{thompson17}. 

An alternative family of imaging algorithms are the ``maximum entropy'' techniques \citep{cornwell85,narayan86,carcamo18} and, more generally, regularized maximum likelihood (\rml) methods \citep{eht-IV-19,nakazato19}. Rather than focusing on image deconvolution, these algorithms instead forward model the visibility measurements using flexible, non-parametric models of the sky brightness distribution conditioned by well-motivated image priors. We will present results of the RML technique \citep[as implemented in the \texttt{MPoL} package;][]{mpol} applied to the MAPS LP in a forthcoming MAPS paper (Czekala et al. \emph{in prep}). 

This article is arranged as follows. In \S\ref{sec:beam} we describe in detail how the configuration of antennas within the interferometric array dictates the characteristics of the synthesized image PSF. We describe how we use the \clean\ algorithm to deconvolve the PSF sidelobe response from synthesized images in \S\ref{sec:cleaning-process}. In \S\ref{sec:JvM} we examine the implications of the so-called ``\citet{jorsater95} effect,'' an important flux scaling issue critical to correctly interpreting faint line emission. We discuss additional strategies of \clean-masking and \uv\ tapering in \S\ref{sec:masking} and \S\ref{sec:taper}, respectively. In \S\ref{sec:signal-to-noise} we discuss how signal to noise ratio might be interpreted in the image products, with reference to the non-imaging matched-filter approach used in \citet{cataldi20} and \citet{ilee20}. We conclude in \S\ref{sec:summary} and describe the available data products from the MAPS LP in \S\ref{sec:data_products}.

\section{Synthesized beams and the impact of \texorpdfstring{$uv$}{uv} coverage} 
\label{sec:beam}
The projected length and orientation of each baseline in the interferometric array directly corresponds to the spatial frequency (\uv\ point) of the visibility function that it measures \citep[Chapter 2;][]{thompson17}. A longer baseline will sample a higher spatial frequency, which corresponds to finer angular scales in the image-plane. As the earth rotates over the course of an observation, the projected length and orientation of a baseline relative to the astronomical source will change and the visibility function will be sampled over a range of \uv\ values \citep[Chapter 5;][]{thompson17}. 

To maximise image sensitivity and resolution, multiple sets of visibilities from short and long baseline array observations are often concatenated together into a single measurement set and imaged. The MAPS program utilized two separate configurations---nominally equivalent to the C43-4 (``short'') and C43-7 (``long'') ALMA configurations---with baselines ranging from 15--1400\,m and 40--3600\,m, respectively. The full details of the array configurations for each observation are summarized in Tables 9 \& 10 of \citet{oberg20}. 

As introduced in \S\ref{sec:intro}, the dirty image is generated by taking the inverse Fourier transform of the visibility measurements while the unsampled spatial frequencies are assumed to carry zero power. Following \citet{briggs95}, the dirty image is formalized as
\begin{equation}
    I_\mathrm{data}(l, m) = C \sum_{k=1}^{N^\dagger} T_k D_k w_k V_{\mathrm{data},k} \exp \left \{ 2 \pi i (u_k l + v_k m) \right \}
    \label{eqn:image}
\end{equation}
where $T_k$ is an optional visibility taper and $D_k$ is a \uv\ density weight. In this instance and in what follows, we assume that the set of visibilities have been augmented to include their complex conjugates, $N^\dagger = 2N$. When making images, it is necessary to sum over the Hermitian conjugates of the visibilities to ensure that the sky image is real. The normalization constant is
\begin{equation}
    C = 1 \Bigg / \sum_{k=1}^{N^\dagger} T_k D_k w_k.
\end{equation}
For the immediate discussion of this section, we assume that $T_k = 1\; \forall k$ and the $D_k$ values correspond to the default density weighting of non-tapered MAPS images, which is \texttt{robust=0.5} \citep{briggs95}. We will return to discuss both tapering and density weighting in more detail in \S\ref{sec:taper}. As discussed in \citet{briggs95}, the units of the dirty image are such that a point source with flux density $S$ will have a peak numerical value of $S$ in the dirty image---for discussion purposes one can reference the flux unit of \jydirtybeam.

The PSF or ``dirty beam'' of the synthesized image is calculated with Equation~\ref{eqn:image} by setting $V_\mathrm{data,k} = 1\,\mathrm{Jy}\; \forall k$, which is the Fourier transform of an impulse sky brightness distribution, i.e., a $1\,\mathrm{Jy}$ point source located at phase center. The PSF is then
\begin{equation}
    B_\mathrm{PSF}(l, m) = C \sum_{k=1}^{N^\dagger} T_k D_k w_k \exp \left \{ 2 \pi i (u_k l + v_k m) \right \}.
    \label{eqn:PSF}
\end{equation}
Phrased differently, if one considers the interferometric array transfer function $W$ under a choice of density weighting and tapering parameters
\begin{equation}
    W(u,v) = \sum_{k=1}^{N^\dagger} T_k D_k w_k \delta(u - u_k, v - v_k),
    \label{eqn:transfer}
\end{equation}
the PSF is its Fourier dual ($B_\mathrm{PSF} \leftrightharpoons W$). The units of \jydirtybeam are technically undefined, since it is not guaranteed that the PSF integrates to a finite volume; however, the maximum is always $B_\mathrm{PSF}(0,0) = 1$. 

\begin{figure*}
\centering
\includegraphics{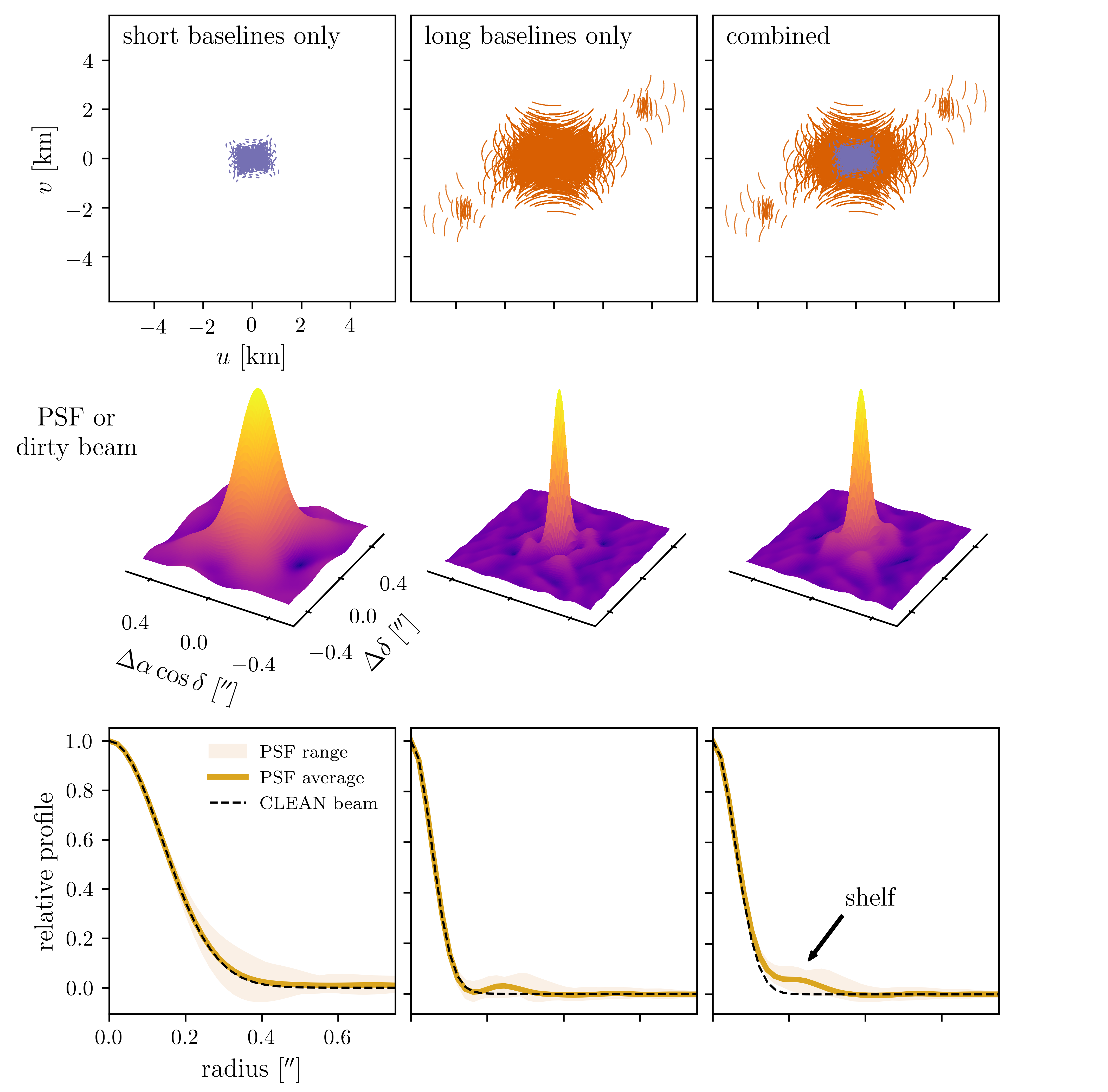}
\caption{How \uv-plane coverage impacts the synthesized point spread function (PSF) or dirty beam. \emph{top row}: The two sets of baselines (i.e., short and long baselines from nominal C43-4 and C43-7 configurations, respectively) resulting from a representative observation of a representative disk in the MAPS LP. \emph{middle row}: The PSF that results from the default Briggs weighting of MAPS ($\texttt{robust}=0.5$). The \clean\ (or restoring) beam is an elliptical Gaussian function fit to the main lobe of the PSF. \emph{bottom row}: The deprojected and azimuthally averaged radial profiles of the PSF and \clean\ beams, with the minimum-to-maximum range of the azimuthal variation of the PSF indicated by the shaded region. Considered individually, ALMA nominal configurations yield dirty beams reasonably approximated (on average) by elliptical Gaussian \clean\ beams. However, the joint baseline configuration results in a PSF with a substantial ``shelf'' to the main Gaussian core.}
\label{fig:baselines}
\end{figure*}

In the first row of Figure~\ref{fig:baselines}, we show the \uv-plane sampling for a representative observation of a disk in the MAPS sample: MWC~480 HCN $J=3-2$ in spectral setting B6-2 \citep[for a full description of the MAPS spectral setup, see Tables 2, 3, \& 4;][]{oberg20}. These samples are split into short and long baselines in the left and middle columns, respectively, and they are combined in the right column. For a sense of scale, the ``combined'' panel contains 408,697 individual visibility samples per spectral channel, which is typical of Band 6 MAPS observations.

The PSFs corresponding to the ``short,'' ``long,'' and ``combined'' baseline sub-selections are shown in the second row of Figure~\ref{fig:baselines}.
Each disk in the MAPS LP was also observed by ALMA at a different elevation, leading to a different set of projected baselines and thus different PSF responses.
For the same disk, all molecular transitions observed at the same time as part of a single spectral setup have the same baseline configuration, but different spectral setups were observed at different stages of array (re-)configuration and therefore have slightly different baseline distributions and PSFs. 

Given the way that ALMA array configurations were originally designed to sample the \uv-plane,\footnote{\url{http://library.nrao.edu/public/memos/alma/main/memo400.pdf} and \url{http://library.nrao.edu/public/memos/alma/main/memo598.pdf}} individual ALMA configurations retain approximately Gaussian beams, with more extended configurations yielding narrower beams and better spatial resolution images (c.f. Figure~\ref{fig:baselines}, columns 1 \& 2). Enhanced resolution comes with tradeoffs, however. 

The maximum recoverable scale (MRS) is a measure of the largest angular scale that can be usefully imaged from a set of visibility measurements. By definition, the more extended configurations of ALMA have fewer short baselines and thus are less sensitive to emission on large spatial scales. If a source has emission on spatial scales larger than the MRS of an array configuration, image flux carried at these low spatial frequencies will not be recovered in the synthesized image \citep[for an analysis of the missing flux using analytic sky brightness distributions, see Appendix A of][]{wilner94}. 

The morphology of molecular line emission in protoplanetary disks is a function of velocity (frequency), with emission near the systemic velocity of the source (transition rest-frame frequency) usually being the most spatially extended. The MRS\footnote{Using the 5th-percentile definition in the ALMA technical handbook \citep{alma_technical_handbook}. See \S2 of \citet{huang20} for a discussion of how the MRS of the combined MAPS configurations affects the interpretation of large-scale CO emission in GM~Aur.} of the nominal C43-7 configuration is only $\approx 1\farcs1$, while all disks targeted in the MAPS LP (IM~Lup, GM~Aur, AS~209, HD~163296 and MWC~480) were known to have emission on spatial scales larger than this \citep[see \S2.1][and references therein]{oberg20}. This makes it necessary to observe with a combination of array configurations to properly sample the visibility function of each MAPS target, especially at the velocities (frequencies) where the emission is the most extended.

The combination of observations from short and long-baseline configurations---at least those utilized by the MAPS LP---yields a dirty beam that has a substantial ``shelf'' at larger radii (Figure~\ref{fig:baselines}, right column). To highlight this shelf, we compare each dirty beam to its corresponding ``\clean'' beam in the third row of Figure~\ref{fig:baselines}. 

The \clean\ beam is an elliptical Gaussian fit to the main lobe of the dirty beam. The beam is reported using the full-width half-maximum along the major and minor axes ($\theta_a$ and $\theta_b$, respectively) and a position angle of the major axis ($\phi$; degrees east of north). The beam power pattern is then
\begin{equation}
    B_\mathrm{\clean}(l,m) = \exp \left(- \frac{1}{2} \left [ \left (\frac{l^\prime}{\sigma_{l^\prime}} \right)^2 + \left( \frac{m^\prime}{\sigma_{m^\prime}} \right )^2 \right ] \right )
\end{equation}
where
\begin{eqnarray}
l^\prime = l \cos \phi - m \sin \phi \\
m^\prime = l \sin \phi + m \cos \phi
\end{eqnarray}
and
\begin{eqnarray}
\sigma_{l^\prime} = \sigma_l = \theta_b / (2 \sqrt{2 \ln 2}) \\
\sigma_{m^\prime} = \sigma_m = \theta_a / (2 \sqrt{2 \ln 2}).
\end{eqnarray}
To convert from units of \jycleanbeam\ to \jyarcsec, for example, one needs to divide by the effective solid angle (angular area) of the \clean\ beam
\begin{equation}
    \Omega_\mathrm{\clean} = \iint B_\mathrm{\clean}(l,m) \; \mathrm{d}l \,\mathrm{d}m = \frac{\pi \theta_a \theta_b}{4 \ln 2}.
    \label{eqn:clean-beam-area}
\end{equation}
This effective solid angle can be calculated by considering the beam response to a spatially uniform source (e.g., the Cosmic Microwave Background). Alternatively, the size and shape of the PSF can be characterized by considering the beam as a three-dimensional solid with its peak normalized to 1. The effective area is then the ``volume'' of this solid in units of $1\times \mathrm{arcsec}^2$, for example, and is graphically illustrated for the dirty beams in the middle row of Figure~\ref{fig:baselines}. The \clean\ beam sizes for all MAPS products are listed in \citet[Table 5]{oberg20}.  To form the one-dimensional beam profiles in the bottom row of Figure~\ref{fig:baselines}, we ``deproject'' both the dirty and \clean\ beam by the aspect ratio of the \clean\ beam and azimuthally average them. In reality, the dirty beams are not azimuthally symmetric, so this is only an approximation for the purposes of visualization (the full range of azimuthal variation is conveyed by the shaded region).

The dirty beam exhibits non-Gaussianity even for the shortest baseline configurations (e.g., a slightly elevated ``tail''). For the combined configurations, the non-Gaussianity manifests as a shelf outside an approximately Gaussian core. Though the shelf may appear small, it is at the root of several issues which ramify throughout the image deconvolution process. We will now describe this process and how we mitigate these issues.

\section{The Cleaning process}
\label{sec:cleaning-process}
\begin{figure*}
    \centering
    \includegraphics{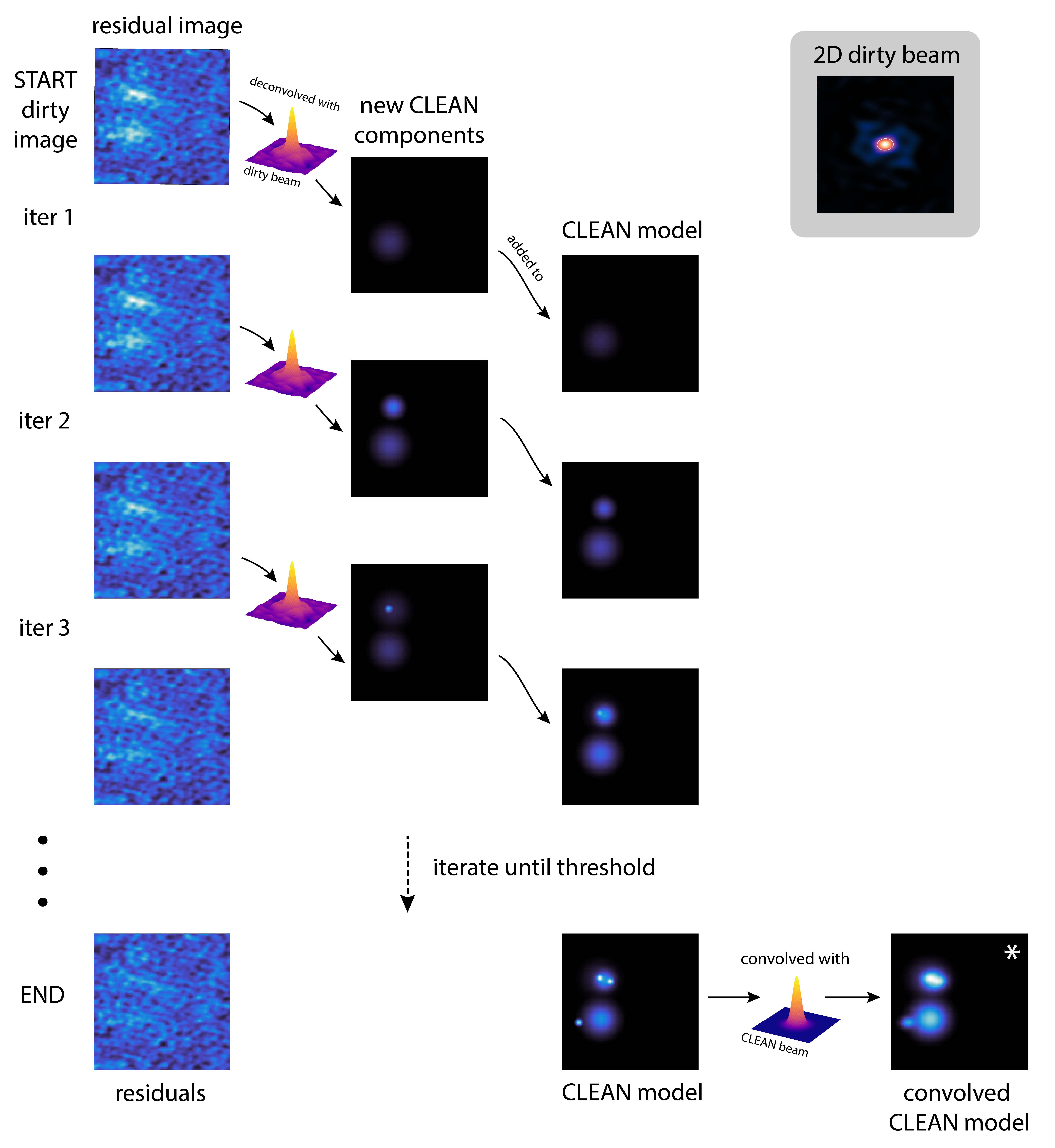}
    \caption{A schematic of the \clean-ing procedure tracking the residual image as new \clean\ components are deconvolved and added to the \clean\ model. The residual image is initialized with the dirty image (which has ``units'' of \jydirtybeam). With each iteration, the residual image is deconvolved by subtracting the convolution of the dirty beam with the new \clean\ component(s), removing the effects of beam sidelobes (a 2D representation of the dirty beam is shown in the upper right grey panel). At the end of the process, the \clean\ model is smoothed by convolution with the \clean\ beam (the FWHM of the \clean\ beam is marked by the white ellipse in 2D dirty beam panel); the resulting image has units of \jycleanbeam. The \clean\ models are shown with an aggressive color stretch ($\sqrt{\;\;}$) to better demonstrate the accumulation of \clean\ components. (*) The same convolved \clean\ model is visualized with a softer linear color stretch in Figure~\ref{fig:workflow}.
    \label{fig:cleaning}}
\end{figure*}

We synthesized and deconvolved image cubes using the \tclean\ task in the Common Astronomy Software Applications (CASA) package version 6.1.0 \citep{mcmullin07}. The salient components of this process are illustrated in Figure~\ref{fig:cleaning} using a single channel containing significant but faint emission from an astrophysical source (AS~209 DCN $J=3-2$). The first algorithmic decision is which functional basis set to use for \clean\ components. The simplest version of \clean\ uses Dirac $\delta$-functions \citep{hogbom74}. We used the more advanced ``multiscale'' algorithm (\texttt{deconvolver="multiscale"}) which is built on a set of variously-sized axisymmetric tapered parabolic components similar to 2D Gaussian functions \citep{cornwell08}. \clean\ components may also take on (small) negative amplitudes so that components placed in later iterations can refine the \clean\ model built from larger positive components placed in earlier iterations. We set the component scales to \texttt{scales=[0, 5, 15, 25]} pixels, where the pixel size was chosen to correspond to $\approx 1/7\mathrm{th}$ of the beamsize. This pixel size adequately oversamples the beam FWHM without making the image size impractically large: the smallest image dimensions were for Band 3 or 0\farcs3 tapered images ($1024\times1024$ pixels) and the largest image dimensions were for Band~6 CO ($2048\times2048$ pixels). 

The \clean\ algorithm starts by setting a ``residual image'' equal to the dirty image and a ``\clean\ model'' equal to a blank image. Each iteration of the \clean\ algorithm introduces new \clean\ components and two things happen. First, the \clean\ model is gradually built up when \clean\ components are placed at locations corresponding to the current peak flux in the residual map. It is possible to use a binary mask to restrict the placement of new \clean\ components. Second, the newly placed \clean\ components are deconvolved from the residual map. The deconvolution is carried out by subtracting the convolution of the \clean\ component with the dirty beam.  The algorithm continues iterating until an exit criterion is triggered. The criterion may be as simple as a maximum number of iterations or it may correspond to a threshold on the noise properties of the residual map. We have chosen the latter and we discuss our choice of masks and thresholds in \S\ref{sec:masking}. 

When the deconvolution procedure has finished, the \clean\ model is convolved with the \clean\ beam to form a convolved \clean\ model image. The final residual image is added to the \clean\ model to form the restored (or ``\clean ed'') image. For multi-channel measurement sets like those of the MAPS LP, \clean\ images and deconvolves each channel with completely independent \clean\ components.

Though \clean\ is best-known for deconvolving dirty beam sidelobes, an arguably more important function of the \clean ing process is creating an interpretable flux model. Because the flux units of the dirty image are technically undefined (\jydirtybeam), any measurements made using a dirty image are highly sensitive to the \uv-sampling distribution. For an extreme example, consider the ill-advised task of measuring the flux of a point source using the dirty image from a two-element interferometer, better known as a fringe pattern \citep[e.g., Ch.\ 9,][]{wilson13}.\footnote{A task more easily carried out by measuring the amplitude of the fringe pattern or (equivalently) using a model-fitting approach (Equation~\ref{eqn:log-likelihood}).} Depending on how one drew their photometric aperture, one could just as easily include or exclude various nulls and maxima of the fringe pattern and measure fluxes ranging from positive peak to negative trough of the sine wave. However, if one were to first deconvolve the fringe PSF from the dirty map, reasonable inferences could be drawn on source flux (if not location, in this contrived example). 

Because the \clean\ model is built up with \clean\ components, it has real, physical interpretabilty: the units of the \clean\ model are \jypixel\ and the units of the convolved \clean\ model are \jycleanbeam. Because the \clean\ beam has finite volume (Equation~\ref{eqn:clean-beam-area}), \clean ed images are on much surer flux-footing than dirty images, whose default dirty beam is not guaranteed to have finite volume.

\clean\ components like $\delta$-functions or tapered parabolic components are rarely a perfect basis to represent a spatially resolved source. However, in the limit of many low-amplitude components, reasonable models of source structure can be achieved. If the \clean\ components were a good match to source morphology, e.g., $\delta$-functions for a field of quasars, then the \clean\ model itself would be a reasonable product to use for analysis. Like the \clean\ model in Figure~\ref{fig:cleaning}, this is rarely achieved in practice; it is more often the case that the higher resolution information that \emph{could} be conveyed by an accurate, native resolution \clean\ model is gladly traded for a more visually pleasing convolved \clean\ model, where errors stemming from a highly discretized but imperfect \clean\ basis set have been low-pass filtered out by \clean\ beam convolution \citep[for a visual explanation, see the discussion of ``optimal resolution'' in][\S5.3]{chael16}. This shortcoming is one reason why flexible, non-parametric imaging techniques can often produce higher resolution image products than \clean\ (see \citealt{eht-IV-19} for a general discussion, and see \citealt{perez19}, \citealt{disk_dynamics20}, or \citealt{jennings20} for discussions specific to protoplanetary disks).  

While the \clean\ procedure may be familiar to many radio astronomers, it is still a non-linear process subject to myriad algorithmic choices like loop gain, threshold stopping criteria, and masking regions, among other parameters embodied in the \tclean\ argument list.\footnote{See the CASA \tclean\ documentation for a full description: \url{https://casa.nrao.edu/casadocs-devel/stable/global-task-list/task_tclean/about} and the NRAO pipeline scripts (\url{https://github.com/AstroChem/MAPS/tree/master/NRAO_processing}) for all parameter settings. Any non-default parameter choices we made are discussed in this article.} When the basis sets are mismatched from the source morphology (e.g., when multi-scale Gaussian components are used to deconvolve a source that is actually composed of concentric rings), it is very important to correctly tune these algorithm parameters to obtain faithfully restored images. It is not common practice within the protoplanetary disk community to publish representations of the \clean\ model in scientific analysis, but we argue that inspection of the \clean\ model (and its potential deficiencies) should be part of any radio astronomy workflow, especially where fine-featured source morphologies are concerned.

\section{The JvM effect and correction}
\label{sec:JvM}

\begin{figure*}
\centering
\includegraphics{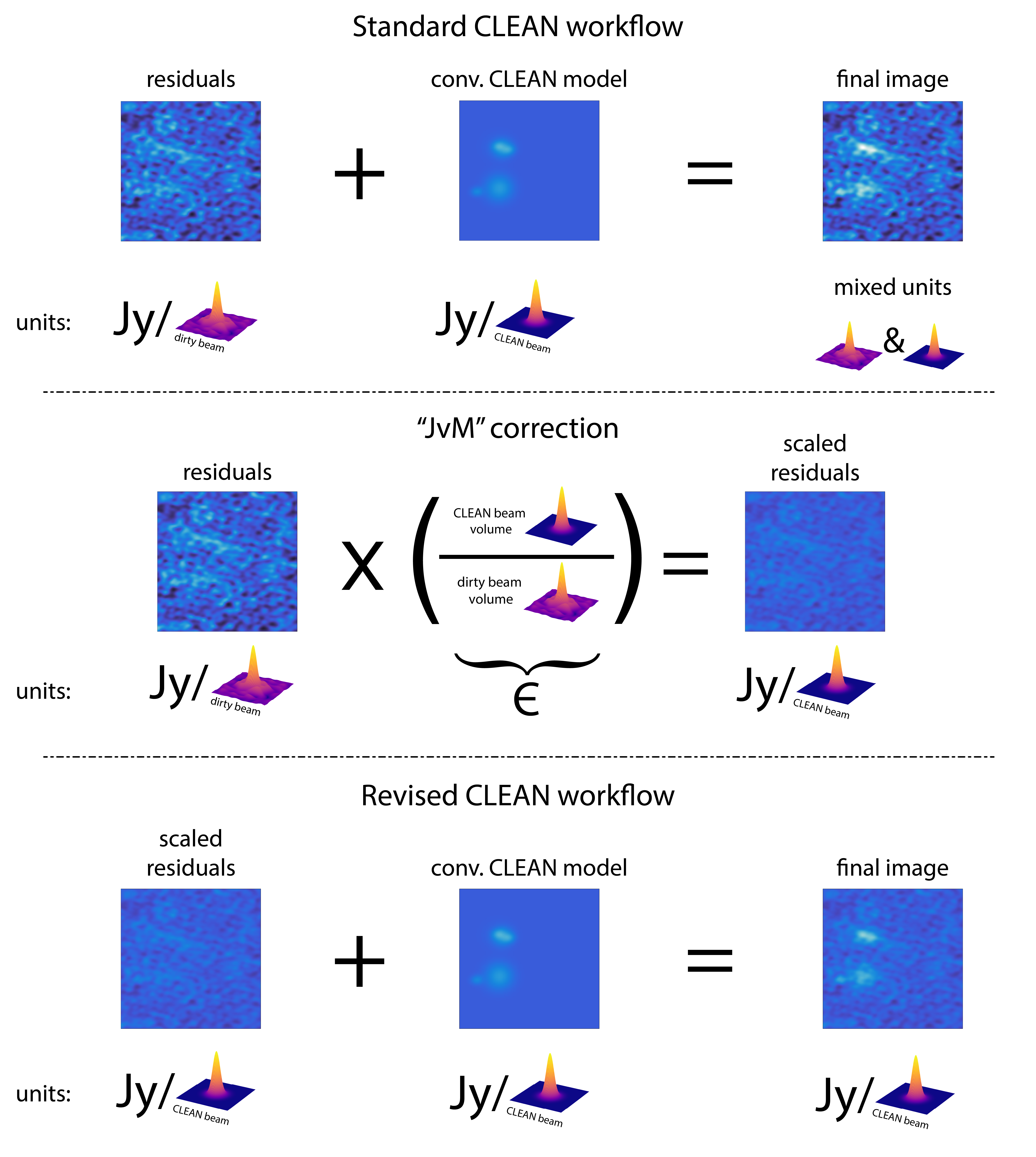}
\caption{After the deconvolution iterations are complete, the residual map is added to the convolved \clean\ model to form the final image. Because the residual map is derived from the dirty image, it has units of \jydirtybeam. In the standard \clean\ workflow (top panel), this creates a final \clean ed image with mismatched units, and therefore compromised interpretability. Following \citet{jorsater95}, one solution is to scale the residual map by the ratio of the \clean\ beam volume to the dirty beam volume, a process that we term the ``JvM correction.'' In the bottom panel we show our revised \clean\ workflow that includes the JvM-corrected residual, which results in a final \clean ed image with the correct intensity units. All images in this figure appear with the same color scale stretch and limits.}
\label{fig:workflow}
\end{figure*}

\subsection{The problem: the \clean\ model and residual image have different units}
Now that we have outlined the broad contours of the \clean ing process, we revisit the final step of the \clean\ algorithm when the final \clean ed image is formed by summing the residual image and the convolved \clean\ model, shown graphically in the top panel of Figure~\ref{fig:workflow} as the ``standard'' workflow. There are usually two reasons why radio astronomers carry out this final step, even though we just discussed reasons why the \clean\ model is a reasonable scientific product on its own terms. One, this gives the final \clean ed image some representation of the thermal noise, which is useful for interpreting the significance of features. Two, the residual map still contains some (ideally small level of) real astrophysical flux that was not adequately deconvolved (e.g., see the ``residuals'' panel in the top row of Figure~\ref{fig:workflow}). Adding the residuals back to the \clean\ model provides some insurance that this real flux at least appears in the final image, though it will still exhibit the effects of the sidelobe response. 

\begin{figure}
\centering
\includegraphics{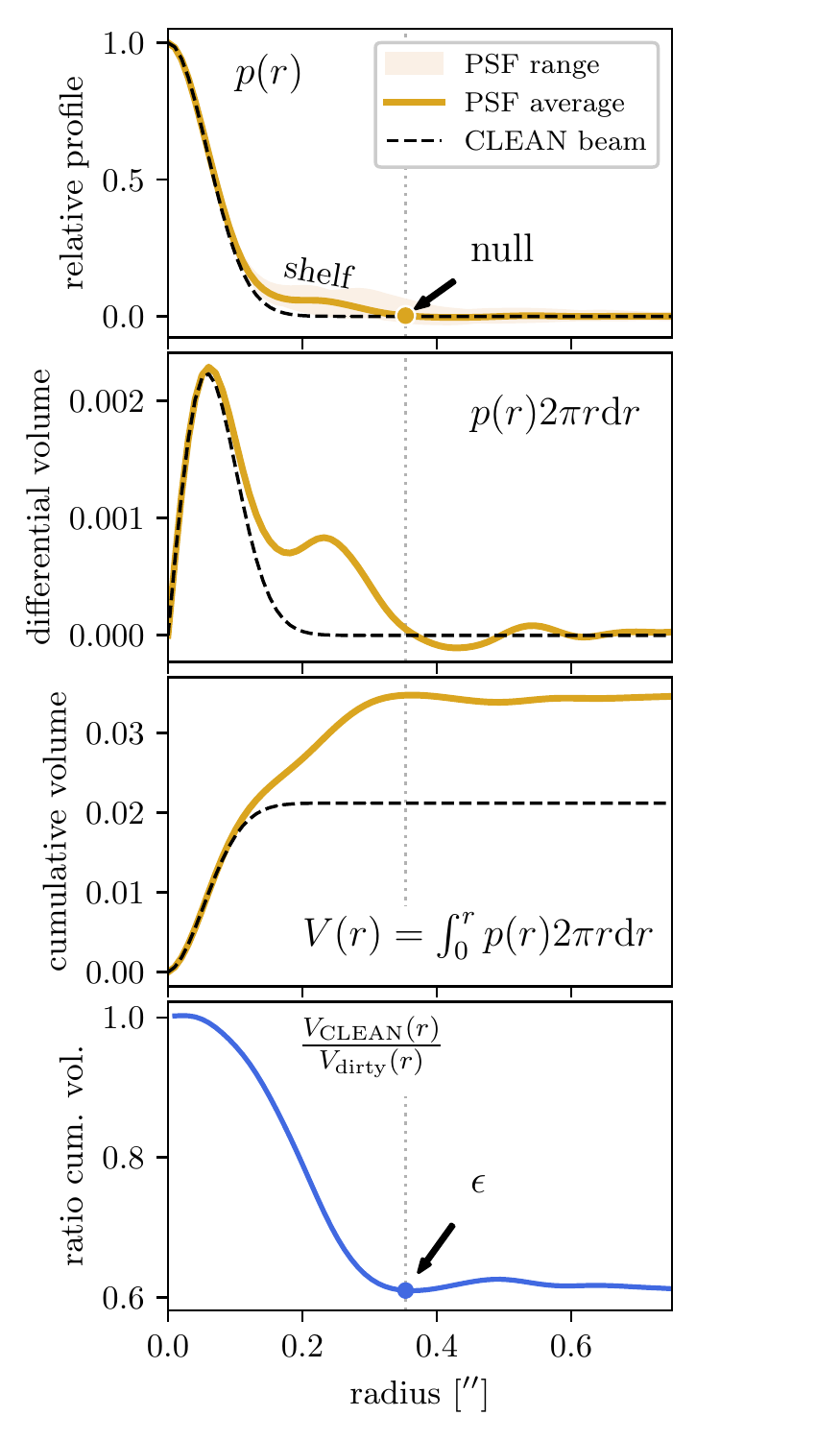}
\caption{How a dirty beam ``shelf'' leads to a divergence of integrated beam volume. \emph{top panel}: the PSF profiles $p(r)$ from the joint baseline configuration in Figure~\ref{fig:baselines}, with the first null (zero-crossing) of the dirty beam labeled as a proxy for the total ``size'' of the dirty beam. \emph{second and third panels}: though the shelf of the dirty beam is small relative to the PSF peak, because it is at large radius, it leads to a significant divergence in the total integrated volume of the dirty beam relative to the \clean\ beam. \emph{last panel}: The ratio of the \clean\ beam volume relative to the dirty beam volume, $\epsilon$, can be used to convert the residuals from the \clean-ing process into the proper units of \jycleanbeam.}
\label{fig:psf-area}
\end{figure}

Because the residual map originated as the dirty image, it technically has units of \jydirtybeam, while the convolved \clean\ model has units of \jycleanbeam. This means that the final \clean ed image is created with mismatched units.  If the \clean\ beam accurately approximates the dirty beam, the unit mismatch is inconsequential. However, if the dirty beam has even a small shelf, such as the $\sim 10\%$ amplitude shelf on the \texttt{robust=0.5} beam shown in Figure~\ref{fig:baselines} and replicated in Figure~\ref{fig:psf-area}, there can be severe implications for accurate flux recovery. Because the shelf occurs at large beam radius, it contributes to a large mismatch in differential volume, even though it is small in relative profile. This small shelf means that a \clean\ beam fit to this dirty beam main lobe will encompass only 60\% of the full dirty beam volume (using the first null as a proxy for dirty beam extent, even though the response extends much further). Therefore, the units \jycleanbeam\ and \jydirtybeam\ differ substantially. The shelf of naturally weighted images is usually even more severe than for robust or uniformly weighted images.\footnote{Because visibility-fitting techniques and forward-model imaging approaches eschew image-plane deconvolution entirely, they have the potential to accurately recover flux even when the dirty beam has a non-trivial shelf.}

To our knowledge, this issue and its implications for flux conservation were first described in \citet[Appendix A]{jorsater95}, and so we term this the ``JvM'' effect throughout the MAPS paper series. We calculate the ratio of beam volumes as
\begin{equation}
    \epsilon = \left . \frac{V_\mathrm{CLEAN}(r)}{V_\mathrm{dirty}(r)} \right |_\mathrm{first\;null},
\end{equation}
using the \clean\ beam and the \texttt{.psf} dirty beam files produced by \tclean.
The procedure originally described in \citet{jorsater95} calculates $\epsilon$ using the ratio of the \clean\ model to the difference between the dirty image and the residual map.\footnote{The ratio of beam volumes need not always be $\epsilon < 1$; depending on the choice of \clean\ beam, which can technically be set to any arbitrary finite-volume beam, values $\epsilon > 1$ are possible. In practice, though, we fit elliptical Gaussian \clean\ beams to the main lobe of the dirty beam and thus $\epsilon \lesssim 1$ across MAPS image products.} Our calculation using the ratio of beam volumes yields a similar result via direct calculation. The bottom panel of Figure~\ref{fig:psf-area} shows a 1D representation of this calculation, though in practice we use the 2D beam profile since the dirty beam is not axisymmetric. 

One might think that the units mismatch can be avoided if only one \emph{fully} \clean ed one's images, i.e., the \clean\ model contained all of the real astrophysical flux and the residual map contained only noise. We concur with \citet{jorsater95} that this is unattainable in any real world application of \clean. Consider an example where the \clean\ threshold is set at $2\times \mathrm{RMS}$. At the end of the \clean ing process the residual map will contain some real astrophysical flux below this threshold while the \clean\ model will contain some components that were erroneously deconvolved from noise spikes above this threshold. Varying the threshold just changes the balance of flux in the residual map or \clean\ model---it is impossible to \clean\ to a zero-flux threshold without also adding a significant number of erroneous components to the \clean\ model. The JvM effect makes imaging \emph{faint} molecular line emission challenging because (by definition) a significant fraction of the flux in each channel will exist at or below a typical \clean ing threshold level (e.g., a few $\times\mathrm{RMS}$). When particularly faint datasets, such as the one chosen to illustrate Figures~\ref{fig:cleaning} \& \ref{fig:workflow} (AS~209 DCN $J=3-2$), are \clean ed to anything but the deepest thresholds (generally inadvisable for other reasons, see \S\ref{sec:masking}), the residuals may contain 50\% or more of the total flux in the image. If one were to erroneously assume the products in the top row of Figure~\ref{fig:workflow} were all in units of \jycleanbeam (as is normal practice), the residuals, \clean\ model, and final image would contain 70\,mJy, 33\,mJy, and 103\,mJy of flux, respectively.

The JvM effect is often less pronounced when synthesizing images of high S/N continuum emission because in that application the \clean\ model will hold a higher proportion of astrophysical flux. However, it may still hamper the characterization of fainter regions within those images.

\subsection{The solution: rescaling the residual image using the ``Jorsater \& van Moorsel (JvM) correction''}
While ``fully \clean ing'' images is not a practical solution to the units mismatch issue, we can approximately convert the residual map from units of \jydirtybeam\ to units of \jycleanbeam. This is achieved by rescaling the residual map by a factor of $\epsilon$ before it is added to the \clean\ model (see Figure~\ref{fig:workflow}, middle panel; for this dataset, $\epsilon=0.359$). We call this the ``JvM correction,'' which we implement using the \texttt{immath} CASA task and the \texttt{.model} and \texttt{.residual} outputs from \tclean. Under the revised workflow, the scaled residuals, \clean\ model and final image in the bottom row of Figure~\ref{fig:workflow} contain 25\,mJy, 33\,mJy, and 58\,mJy of flux, respectively (resulting in a dramatic ~50\% change in total flux compared to the standard workflow). As demonstrated by this faint dataset, failure to apply the JvM correction can have a profound effect on both the total flux contained within an image and the morphological characteristics of that flux (c.f. Figure~\ref{fig:workflow} final images).

The $\epsilon$ values for all MAPS image cubes are listed in Table 11 of \citet{oberg20}. Band 3 image cubes typically have $\epsilon$ values in the range 0.7 - 1.0, while band 6 image cubes typically have $\epsilon$ values in the range 0.2 - 0.7. The effect of the JvM correction becomes more significant the more $\epsilon$ deviates from 1.

\paragraph{JvM still matters for signal-free channels!} 
Since the JvM correction modifies the residual map, this creates ambiguity around an image-plane RMS measurement. Technically, even signal-free channels with zero \clean\ components still need to be corrected for the JvM effect since they are formed from a residual map in units of \jydirtybeam\ but are reported in units of \jycleanbeam. Unless otherwise stated, throughout the MAPS paper series we specify RMS values in \jycleanbeam\ measured from emission-free images corrected for the JvM effect (but not yet corrected for the primary beam sensitivity).\footnote{Thus, the RMS values are equivalent to an RMS value measured at the phase center of an image corrected for the primary beam sensitivity.} 

A notable exception is when we specify \clean ing thresholds. Since the \tclean\ procedure references threshold values in the dirty map itself, we report these thresholds with respect to RMS values measured in the non-JvM-corrected dirty map. 

\section{Masking to improve the \clean\ model}
\label{sec:masking}
Binary image plane masks are useful to guide the placement of \clean\ components to locations believed to correspond to physical flux. Simple elliptical masks are often reasonable choices, especially for single-channel continuum images. Since the spatial distribution of molecular line emission in a protoplanetary disk changes considerably (but predictably) as a function of observing frequency \citep[corresponding to velocity, e.g., Figure 3,][]{oberg20}, using the same elliptical mask for all channels means that large areas of blank sky will be at risk of having erroneous \clean\ components placed within them. 

CASA allows masks to be hand-drawn for complicated spatial emission. This is often the most accurate option, but mask drawing quickly becomes onerous for even a single large spectral cube. For the MAPS LP program, the time investment is prohibitive. CASA's  \texttt{auto-multithresh} algorithm \citep{kepley20} is a promising solution, however, given its computational overhead, we instead exploited the Keplerian rotation pattern of the protoplanetary disk \citep[e.g.,][]{horne86,semenov08} to create parametric \clean\ masks \citep[e.g.,][]{rosenfeld13}.

\begin{figure*}[t]
\centering
\includegraphics{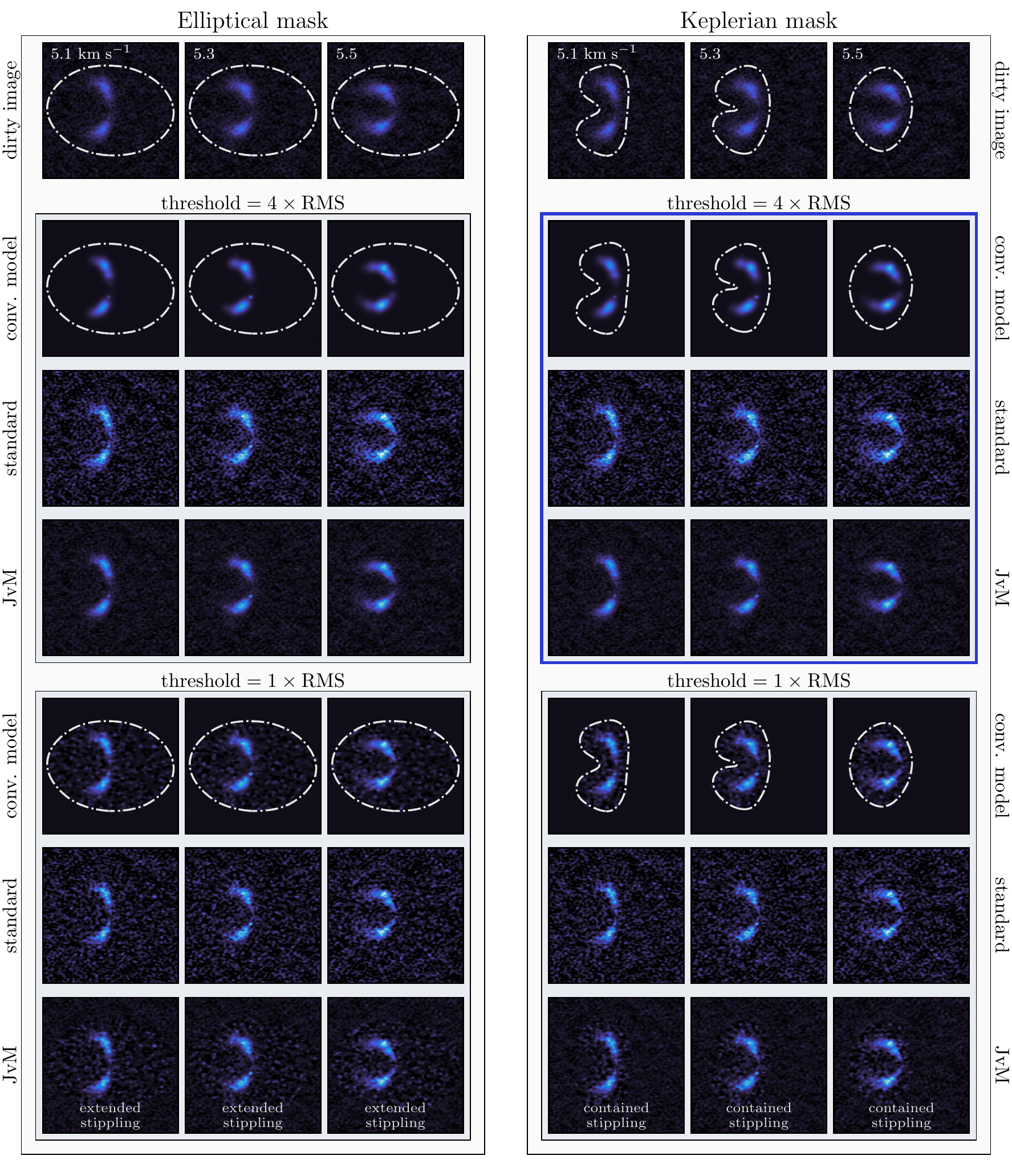}
\caption{A comparison of elliptical (left column) vs. Keplerian (right column) binary masks for three channels of the AS~209 ${}^{13}\mathrm{CO}$ $J=2-1$ measurement set \clean ed to an ideal depth of $4\times\mathrm{RMS}$ (top grouping) and over-\clean ed to a deeper yet-still-plausible depth of $1\times\mathrm{RMS}$ (bottom grouping). All panels are on the same color scale stretch. If the parameters of the \tclean\ algorithm are set optimally, a well-designed mask will not significantly improve image quality over a generic mask, as demonstrated by the identical images generated using the $4\times\mathrm{RMS}$ threshold. However, if an image is \clean ed to a deep noise threshold, a poorly designed mask will more easily allow components to be erroneously added to the \clean\ model. These errant components are barely noticeable in the ``standard'' \clean\ workflow (which unfortunately has inconsistent flux units, Figure~\ref{fig:workflow}), but when the JvM correction scales the residuals by a factor of $\epsilon$ the stippled pattern becomes apparent. Well-designed Keplerian masks can mitigate but not eliminate this behavior. The default choices for MAPS data products are labeled with a thick blue border (Keplerian masks, 
\texttt{threshold=}$4\times\mathrm{RMS}$).}
\label{fig:masking}
\end{figure*}

\subsection{Keplerian masking}
\label{subsec:keplerian_masks}
We generated a Keplerian mask for each disk by the following procedure. Each image-plane pixel $(l, m)$ was first deprojected into disk-centric cylindrical coordinates $(r, \phi, z)_\mathrm{disk}$, based on an assumed disk inclination $i$, position angle PA, and constant emission surface slope \citep[$z/r$, e.g.,][]{Teague2019a} representing the surface of an optically thick cone in three dimensions. Disk inclination and position angle values were drawn from the literature \citep[see][Table 1]{oberg20}, while emission surface slope was refined by hand as described below. The disk-centric coordinates correspond to the location where a ray drawn from pixel $(l,m)$ intersects the surface of the cone. Depending on the disk inclination and orientation, not all pixels necessarily intersect the cone. Using the disk-centric coordinates for each pixel, the projected Keplerian velocity is
\begin{equation}
    v_{\rm Kep,\, proj}(r,\, z) = \sqrt{\frac{G M_\star r^2}{(r^2 + z^2)^{3/2}}} \sin(i) \cos(\phi)
\end{equation}
where $M_\star$ is the central stellar mass. The total projected velocity component at each pixel is the sum of the Keplerian rotation and the systemic velocity, $v_0 = v_{\rm Kep,\, proj} + v_{\rm LSR}$.

Motivated by the fact that disk temperature declines with increasing $r$, the intrinsic line width is also assumed to narrow following a power law profile
\begin{equation}
    \Delta V(r) = \Delta V_0 \times \left( \frac{r}{1\arcsec} \right)^{\Delta V_q}
\end{equation}
where $\Delta V_q \leq 0$. An initial mask was generated such that for a channel with central velocity $v_{\rm chan}$ and width $\Delta v_{\rm chan}$, masked regions satisfied $|v_{\rm chan} - v_0| \leq (\Delta v_{\rm chan} + \Delta V)$. For image cubes with low spectral resolution, the channel spacing $\Delta v_\mathrm{chan}$ dominates the mask. For image cubes with high spectral resolution, the line width $\Delta V$ is of greater importance in determining masked regions. After the initial mask was generated, it was convolved with a 2D Gaussian profile to smooth the mask edges (with FWHM $\theta_{\rm conv}$ tuned to provide an adequate ``buffer'' region near otherwise sharp contours). The full script used to generate the masks is available in the \texttt{keplerian\_mask} package \citep{teague_kepmask}. 

We tuned the mask parameters to match the $^{13}$CO $J=2-1$ emission for each disk; the final parameters are listed in Table~\ref{tab:mask_parameters_13CO}. The mask parameters should not be considered true properties of the protoplanetary system, rather they are the parameters that best represent the spatial distribution of emission using the framework described above. The same mask parameters are used for all default image products for all observed transitions, with the exception of $^{12}$CO $J=2-1$, whose extended and complex emission morphology warranted bespoke, hand-drawn masks. For some specialized applications, it was advantageous to tune the Keplerian mask parameters from those values used here to produce the default image products. Where applicable, these alternative masks are noted in MAPS publications.

\begin{deluxetable}{lccccc}
\tablecaption{Keplerian Mask Parameters \label{tab:mask_parameters_13CO}}
\tablehead{
\colhead{Source} & \colhead{$r_{\rm out}$} & \colhead{$\Delta V_0$} & \colhead{$\Delta V_q$} & \colhead{$z \, / \, r$} & \colhead{$\theta_{\rm conv}$} \\
\colhead{} & \colhead{\footnotesize($\arcsec$)} & \colhead{\footnotesize(${\rm m\,s^{-1}}$)} & \colhead{} & \colhead{} & \colhead{\footnotesize($\arcsec$)}}
\decimals
\startdata
IM~Lup      & 5.8 & 400 & -1.0 & 0.3 & 0.5 \\
AS~209      & 1.8 & 400 & -0.5 & 0.1 & 0.5\\
GM~Aur      & 3.0 & 500 & -0.5 & 0.2 & 0.3 \\
HD~163296   & 5.6 & 500 & -1.0 & 0.3 & 0.3 \\
MWC~480     & 3.0 & 300 & -0.5 & 0.3 & 0.3 \\
\enddata
\end{deluxetable}

\subsection{``Stippling'' with deeply \clean ed models}
\label{subsec:stippling}

In Figure~\ref{fig:masking} we compare the results of the \clean ing process for three channels of an example measurement set (AS~209 ${}^{13}\mathrm{CO}$ $J=2-1$) using an elliptical mask vs. a Keplerian mask. To demonstrate the impact of the differently shaped masks, we \clean ed to two different depths: $4\times\mathrm{RMS}$ and $1\times\mathrm{RMS}$. As noted previously, for \clean\ applications we used the RMS measured from a signal-free region prior to the JvM correction.

If the measurement set is properly calibrated and the \tclean\ parameters are set optimally, i.e., the basis set is an adequate match to the source morphology, loop gain is sufficiently conservative given the complexity of the deconvolution task, and the \clean ing threshold is appropriate, a well-designed mask will not significantly improve the end result (cf. elliptical vs. Keplerian masks at a threshold =  $4\times\mathrm{RMS}$ in Figure~\ref{fig:masking}). If the algorithm is tuned correctly, \clean\ will eventually converge as it steadily but surely deconvolves real astrophysical flux (presumed to be responsible for the brightest features in the residual map at each iteration) down to the threshold. Through careful experimentation, we arrived at $4\times\mathrm{RMS}$ as the ideal cleaning depth to balance the number of erroneous \clean\ components against flux retained in the residual map.

However, the optimal \tclean\ parameters can be difficult to ascertain without some trial and error, and the threshold is an example of a parameter that is sometimes difficult to tune \emph{a priori}. In these instances, properly-fitting binary masks can guard against the incorrect placement of \clean\ components and mitigate (but not eliminate) some of the adverse effects of improperly set \tclean\ parameters. In the \texttt{threshold=}$1\times\mathrm{RMS}$ example in Figure~\ref{fig:masking}, we see that numerous low-amplitude \clean\ components are added to the model, corresponding to noise spikes above the threshold level  that were erroneously deconvolved from the residual map (top row: ``conv.\ model''). Because the elliptical mask encompasses a larger area without plausible astrophysical flux, there are more opportunities for components to be incorrectly deconvolved.

Under the standard \clean\ workflow, the visual appearance of the final product (labeled ``standard'' in Figure~\ref{fig:masking}) is not significantly affected at either threshold. However, we remind the reader that the flux units of the \clean ed image produced under the standard workflow are inconsistent (\S\ref{sec:JvM}), and therefore any residual flux that was not deconvolved (particularly that outside the mask) is not represented with the appropriate strength. Moreover, as a matter of principle, it is not desirable to have a \clean\ model contain components known to correspond to pure noise, especially if it were to be used for further data reduction or analysis \citep[e.g., self-calibration;][]{brogan18}.

When the residual map and \clean\ model are properly combined under the JvM workflow (labeled ``JvM'' in Figure~\ref{fig:masking}), the defects of the \clean\ model become apparent in the form of a stippled pattern. The reason this stippling occurs is because when $\epsilon < 1$, the flux distribution loses support under the \clean\ model: the \clean\ algorithm deconvolves the residual map with a dirty beam that is substantially larger than the \clean\ beam can restore. When the residual map is properly scaled to \jycleanbeam\ using the JvM correction, the previously artificially high residual map drops to a lower level and the stippling pattern appears. The stippling effect is most apparent in regions of low astrophysical flux and might lead to the misinterpretation of substructure in what would otherwise be interpreted as smooth regions \citep{disk_dynamics20}. As mentioned, we used the carefully tuned $4\times\mathrm{RMS}$ cleaning depth to minimize the appearance of stippling across the MAPS data products (see the panels with a blue border in Figure~\ref{fig:masking}).

\section{UV tapering to improve beam shape}
\label{sec:taper}
The primary goal of this section is to describe how the tapering coefficients $T_k$ of the dirty image equation (Equation~\ref{eqn:image}) can be modified to standardize PSF characteristics throughout the MAPS LP and aid comparisons between molecules and across disks. Because PSF tapering is interrelated with the density weighting coefficients $D_k$, however, we first briefly review how uniform, natural, and robust weighting affect PSF shape through \uv\ cell averaging and density weighting \citep[see also Chapter 10.2.2;][]{thompson17}. We also comment on how the images generated with these weighting schemes relate back to the data likelihood originally discussed in \S\ref{sec:intro}. Then, we discuss how the tapering and density weighting coefficients can be co-varied to achieve a range of desired PSFs. In Figure~\ref{fig:tapering} we demonstrate how the density weighting schemes and tapering coefficients affect the beam sidelobes, beam (non)axisymmetry, and JvM correction factor.

\subsection{Cell averaging, density weighting, and maximum likelihood images}
As discussed in the introduction (\S\ref{sec:intro}), Equation~\ref{eqn:image} is used to synthesize a ``dirty image'' from a set of discrete visibility samples $\{V_{\mathrm{data},k}\}^N_{k=1}$. In practice, the visibilities and their Hermitian conjugates are first ``gridded,'' or interpolated, onto a regularly spaced array of $u_i$, $v_j$ points\footnote{The method by which visibilities are interpolated to grid center points turns out to have implications for image fidelity, especially for high dynamic range applications. Nearest-neighbor or linear interpolation can result in significant numerical errors; convolutional regridding schemes using prolate spheroidal wave functions are good choices to minimize these errors \citep{schwab84,czekala15a}. For the discussion that follows, however, the main conclusions are unchanged if simple nearest neighbor interpolation is assumed.} so that the fast Fourier transform may be used to accelerate the imaging calculation. This grid presents a convenient partition centered on the $u_i$,$v_j$ pairs with cell sizes $\Delta u$ and $\Delta v$ with which to a) average multiple visibilities contained within the cell and b) use as a bounding box to calculate local sample density. The cell sizes used for averaging and density calculations need not be the same as those in the gridded array, but for demonstration purposes we assume that they are.

For the sake of discussion, we consider an arbitrary cell $i,j$ containing $L$ visibilities, indexed such that each $l$ index corresponds to some specific $k$ index. For example, $l= \{1, 2, 3, \ldots,L\}$ might correspond to $k = \{304, 305, 307, \ldots, 320\}$, depending on how exactly the visibilities are indexed relative to the grid boundaries. 
When all $L$ visibilities inside the cell are gridded, the ``effective'' visibility located at the cell center is
\begin{equation}
    V_{\mathrm{grid},i,j} = \sum_{l=1}^L T_k D_k w_k V_{\mathrm{data},k}.
\end{equation}

To simplify the comparison of density weights that follows, we will assume that there is no tapering ($T_k = 1$) and that all visibilities within a cell have a common expectation value $\langle V_{\mathrm{data},k} \rangle_{i,j} = {\cal V}(u_i, v_j)$, since the cell size is assumed to be small relative to the expected features in the visibility function (i.e., the chosen image size and resolution ``Nyquist sample'' the visibility function).

\begin{figure*}[t]
\centering
\includegraphics{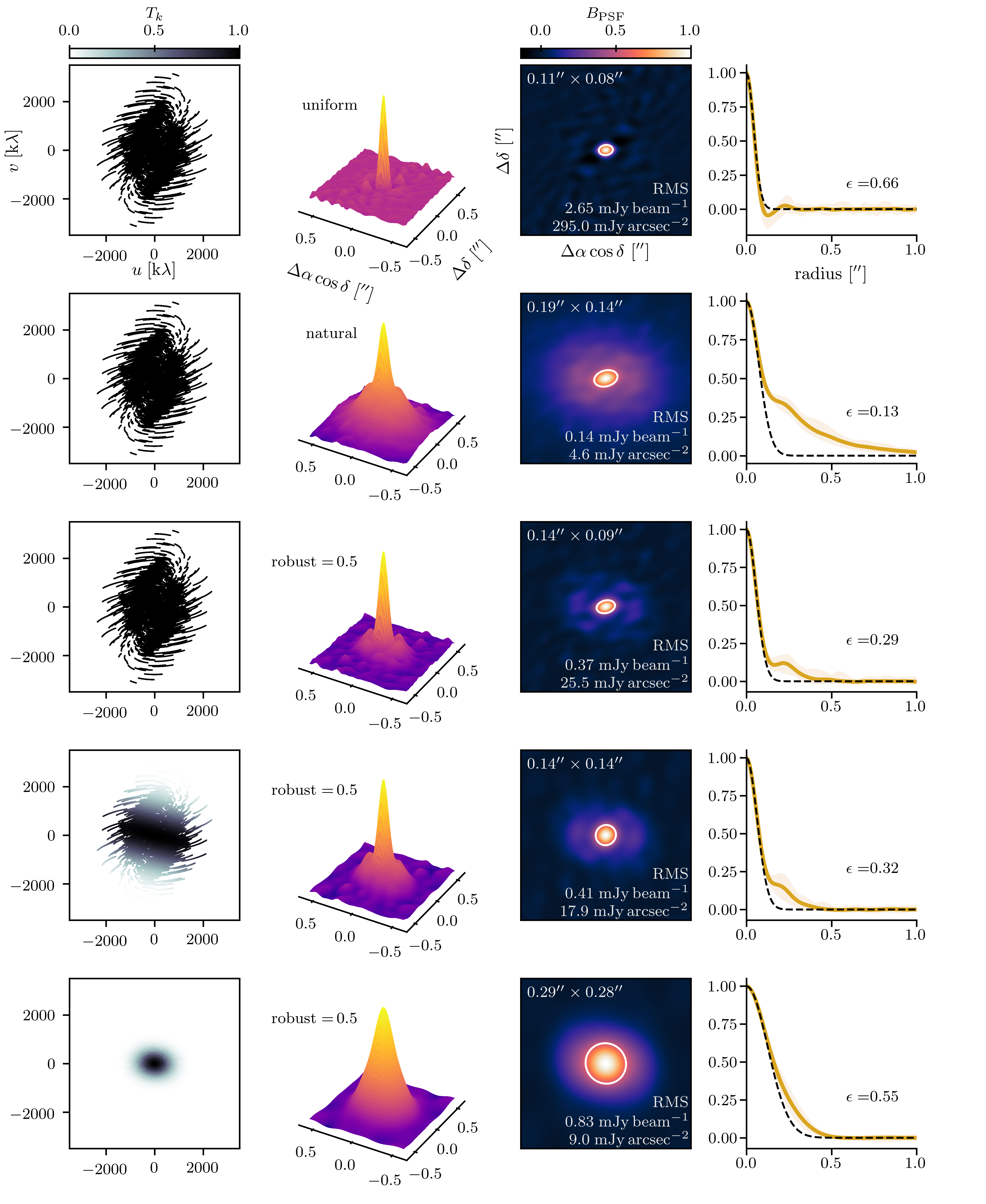}
\caption{A representation of how the visibility tapering profile $T(u,v)$ (left column) and density weighting scheme influence the dirty beam resolution, point source sensitivity, PSF profile and JvM correction factor $\epsilon$. The same dirty beam profile is shown in the second, third, and fourth columns: the three dimensional view best represents beam structure, the two dimensional view best conveys beam ellipticity (white ellipse represents FWHM), and the one dimensional view best represents the divergence from a Gaussian \clean\ beam. Each row represents a different combination of tapering profile and robust value. The first three rows are untapered, while the final two rows are tapered to circular FWHM resolutions of $0\farcs15$ and $0\farcs30$, respectively (the final $\leq 5\%$ adjustment is carried out using the \texttt{imsmooth} task, which is not shown). }
\label{fig:tapering}
\end{figure*}

\paragraph{Uniform weighting}
``Uniform'' weighting is arrived at by minimizing the sidelobe level of the main beam \citep[for a derivation, see \S3.3.3;][]{briggs95}, which yields constant density weights for all visibilities within the cell
\begin{equation}
    D_k = \frac{1}{w_{\mathrm{cell},i,j}}
\end{equation}
where $w_{\mathrm{cell},i,j} =  \sum_{l=1}^L w_l$. The effective value of the uniformly gridded visibility is 
\begin{equation}
    V_{\mathrm{grid},i,j} = \frac{\sum_{l=1}^L w_l V_{\mathrm{data},l}}{\sum_{l=1}^L w_l}  = \langle V_{\mathrm{data},k} \rangle_{i,j}.
    \label{eqn:uniform-effective}
\end{equation}
Uniform weighting delivers the minimum variance estimate of the visibility mean, since it is obtained by weighting each sample by its inverse variance ($w_k = 1/\sigma_k^2$). Moreover, the statistical uncertainty on $V_{\mathrm{grid},i,j}$ is encapsulated with $w_{\mathrm{cell},i,j}$ in the same way as for ungridded visibilities $V_{\mathrm{data},k}$ with $w_k$. While uniform weighting usually produces the highest resolution dirty beam (Figure~\ref{fig:tapering}, first row), a downside is that the dirty image usually has the worst sensitivity.

\paragraph{Natural weighting}
The RMS thermal noise in the dirty image is minimized when
\begin{equation}
    D_k = 1,
\end{equation}
since $w_k = 1/\sigma_k^2$ \citep[for a derivation, see \S3.3.1;][]{briggs95}. This is called ``natural'' weighting and results in an image that has the lowest RMS and thus greatest sensitivity. The effective value of the naturally gridded visibility is
\begin{equation}
    V_{\mathrm{grid},i,j} = \sum_{l=1}^L w_l V_{\mathrm{data},l}. 
    \label{eqn:natural-effective}
\end{equation}
Natural weighting diverges from estimating $\langle V_{\mathrm{data},k} \rangle_{i,j}$ and instead upweights those grid cells containing large quantities of high S/N visibilities. Natural weighting results in a beam (Figure~\ref{fig:tapering}, second row) that is useful for detection (especially of point sources), but at the cost of substantial sidelobes negatively impacting resolution.

\paragraph{Robust weighting}
\citet{briggs95} developed the ``robust'' weighting scheme 
\begin{equation}
    D_k = \frac{1}{1 + w_{\mathrm{cell},i,j} f^2}
\end{equation}
where
\begin{equation}
    f^2 = (5 \times 10^{-R})^2 \Bigg / \left (\frac{\sum_i \sum_j w_{\mathrm{cell},i,j}^2 }{\sum^{N^\dagger}_{k=1} w_k} \right).
\end{equation}
Values of $R = 2$ approximate natural weighting (maximizing point source sensitivity) while values of $R = -2$ approximate uniform weighting (maximizing resolution). Intermediate $R$ values offer a tradeoff between these two extremes. Because the performance curve is nonlinear \citep[Figure 3.23,][]{briggs95}, significant resolution gains can be achieved at minimal loss in \emph{point source} sensitivity (e.g., $\mathrm{mJy}$), though loss in surface brightness sensitivity is more significant (see Figure~\ref{fig:tapering}, third column), especially when expressed in units with constant solid angle (e.g., $\mathrm{mJy}\,\mathrm{arcsec}^{-2}$ or brightness temperature [$K$]).

The effective value of the robustly gridded visibility is 
\begin{equation}
    V_{\mathrm{grid},i,j} = \sum_{l=1}^L \frac{w_l V_{\mathrm{data},l}}{1 + w_{\mathrm{cell},i,j} f^2} \Bigg / \sum_{k=1}^{N^\dagger} \frac{w_k}{1 + w_{\mathrm{cell},i,j} f^2}
    \label{eqn:robust-effective}
\end{equation} 

\paragraph{Maximum likelihood images}
If the visibility function ${\cal V}$ were sampled at all $u,v$ locations where it had significant power, then $V_\mathrm{grid}$ would have a direct relationship to the sky brightness via the inverse Fourier transform. Unfortunately, this is never achieved for actual sub-mm interferometric observations of astrophysical sources, and the impact of the transfer function ($W(u,v)$, Equation~\ref{eqn:transfer}) must be considered when producing images from visibility samples.

Nevertheless, because visibility model fitting (\S\ref{sec:intro}) is a common and highly useful procedure for parameter inference---especially for annular protoplanetary disk structures---it is worthwhile to consider how uniform, natural, and robust images relate back to ${\cal V}$. By referencing the effective visibility under each weighting scheme (Equations~\ref{eqn:uniform-effective}, \ref{eqn:natural-effective}, \& \ref{eqn:robust-effective}), we see that only uniform weighting converges to the expectation value of the visibility function in that cell \citep{briggs95}, while natural and robust weighting instead result in an effective visibility that is tied to the number and quality of visibility measurements within that cell. Therefore if the dirty image were used as a sky plane model $I(l,m)$, only the uniform dirty image would maximize the data likelihood function (Equation~\ref{eqn:log-likelihood}).

Of course, there are legitimate reasons to use other weighting schemes, such as the need to maximize sensitivity with natural weighting for a detection experiment. And in practice, given its non-linear tradeoff curve, the most scientifically useful weighting scheme for the detection and characterization of faint protoplanetary disk features is often an intermediate robust value, like the \texttt{robust=0.5} we used as the default setting for untapered MAPS products.

\paragraph{The \clean ing process and maximum likelihood}
The dirty image is constructed under the assumption that all unsampled spatial frequency components are set to zero power. As the \clean ing process deconvolves the dirty image and image plane components are added to the \clean\ model, the corresponding visibility plane model $V_\mathrm{model}$ is effectively interpolated from the gridded locations to otherwise unsampled ($u,v$) values. Because the image plane representation is directly equivalent to the inverse Fourier transform of the visibility plane model, the accuracy of the interpolated values relative to truth mirrors the quality of the deconvolution process. Using a basis set appropriate to the source morphology (e.g., choosing multi-scale \clean\ over H\"{o}gbom \clean\ for extended emission) can result in more accurate Fourier interpolations/image reconstructions, particularly at higher spatial frequencies/finer resolutions.

Each iteration of the \clean\ algorithm also brings $V_\mathrm{model}$ closer to maximizing the likelihood function \citep{schwarz78}. However, $V_\mathrm{model}$ will only achieve the maximum possible likelihood value when $V_{\mathrm{model},k} = V_{\mathrm{data},k} \;\forall k \in 1,\ldots,N$, which can happen in the absence of noise or in the case that the entire image is fully cleaned to a $0\,\mathrm{Jy}$ threshold (the Fourier equivalent of overfitting). Most practical applications will stop \clean ing when a noise threshold is reached in the residual map. 

Because imaging is an ill-defined inverse process, there exist an infinite number of maximum (or approximately maximum) likelihood images (and therefore models) consistent with the data: these have $V_{\mathrm{model},k} \approx V_{\mathrm{data},k} \;\forall k \in 1,\ldots,N$ but may take on any value of $V_{\mathrm{model}}$ at unsampled spatial frequencies $u,v$. Because beam characteristics will affect the order in and scales at which flux is deconvolved from the dirty image, for any spatially resolved source the deconvolution algorithm will likely converge to different \clean\ models under different visibility weighting schemes.

\subsection{Tapering}
Visibility tapering profiles can be used to ``force'' beams of arbitrary dimensions \citep{briggs95}. For MAPS, we produced fiducial data products using tapered visibilities to boost image-plane sensitivity to low surface brightness features (at the expense of resolution) and provide a common circular beam that is useful for comparing molecular species across the Band 3 and Band 6 observations \citep[for a listing of all fiducial beam sizes, see Table 5 of][]{oberg20}.

Beam tapers can be described by a multiplicative profile in the \uv-plane $T(u,v)$ or equivalently by convolution with a kernel in the image-plane $t(l,m)$. If we wish to taper our original dirty beam, whose main lobe is approximately an elliptical Gaussian, to a dirty beam whose main lobe is approximately a circular Gaussian, the appropriate functional form of the taper is an elliptical Gaussian. In analogy with the \clean\ beam (\S\ref{sec:beam}), this profile is described by a position angle ($\phi_T$) and beam dimensions. In the \uv-plane, these FWHMs are $\Theta_a$, $\Theta_b$ in units of k$\lambda$. In the image plane, these are $\theta_a$, $\theta_b$, in units of arcsec. The Gaussian FWHMs are related\footnote{We derived the FWHM relationships following the Fourier dual relationships for Gaussians, which are more simply phrased in terms of $\sigma$ values: $\sigma_{u,v} = 1/(2 \pi \sigma_{l,m})$ \citep[e.g.,][]{bracewell00}. It appears there is an inconsistency in the CASA 6.1 implementation of \texttt{uvtaper} in \tclean\ where the incorrect FWHM relationship 
\begin{equation}
    \Theta =  10^{-3} \left ( \frac{60^2 \times 180\,\mathrm{arcsec}}{\pi\,\mathrm{radians}} \right )  \frac{1}{\theta} 
\end{equation}
is used to calculate the \texttt{uvtaper} profile, which equates to a scale factor difference of $\approx 1.13$. We emulated this behaviour such that the final images were tapered to the correct target resolution.}
such that 
\begin{eqnarray}
\Theta_{a} &=&  10^{-3} \left ( \frac{60^2 \times 180\,\mathrm{arcsec}}{\pi\,\mathrm{radians}} \right )  \frac{4 \ln 2}{\pi \theta_a} \\
\Theta_{b} &=& 10^{-3} \left ( \frac{60^2 \times 180\,\mathrm{arcsec}}{\pi\,\mathrm{radians}} \right ) \frac{4 \ln 2}{\pi \theta_b}.
\end{eqnarray}
We follow a convention such that $\Theta_a > \Theta_b$, which implies that $\theta_a < \theta_b$. The multiplicative \uv-plane profile is 
\begin{equation}
    T(u,v) = \exp \left(- \frac{1}{2} \left [ \left (\frac{u^\prime}{\sigma_{u^\prime}} \right)^2 + \left( \frac{v^\prime}{\sigma_{v^\prime}} \right )^2 \right ] \right )
\end{equation}
with
\begin{eqnarray}
u^\prime = u \cos \phi_T - v \sin \phi_T \\
v^\prime = u \sin \phi_T + v \cos \phi_T
\end{eqnarray}
and
\begin{eqnarray}
\sigma_{u\prime} = \sigma_u = \Theta_b / (2 \sqrt{2 \ln 2}) \\
\sigma_{v^\prime} = \sigma_v = \Theta_a / (2 \sqrt{2 \ln 2}).
\end{eqnarray}
In the image-plane, the convolutional profile is
\begin{equation}
    t(l,m) = \exp \left(- \frac{1}{2} \left [ \left (\frac{l^\prime}{\sigma_{l^\prime}} \right)^2 + \left( \frac{m^\prime}{\sigma_{m^\prime}} \right )^2 \right ] \right )
\end{equation}
with
\begin{eqnarray}
l^\prime &=& l \cos \phi_T - l \sin \phi_T \\
m^\prime &=& m \sin \phi_T + m \cos \phi_T
\end{eqnarray}
and
\begin{eqnarray}
\sigma_{l\prime} &=& \sigma_l = \theta_b / (2 \sqrt{2 \ln 2}) \\
\sigma_{m^\prime} &=& \sigma_m = \theta_a / (2 \sqrt{2 \ln 2}).
\end{eqnarray}
Both the \uv\ tapering and image convolutional profiles are normalized to peak values of 1. Although it is possible to analytically calculate tapering parameters given a desired beam size \citep[Appendix C of][]{briggs95}, we found that there is considerable ``mechanical backlash'' in CASA v6.1.0's beam fitting subroutines due to the pixelized representation of the dirty beam. This means that a smooth, monotonic relationship between tapered dirty beam size and fitted \clean\ beam size does not exist; in practice we found that direct forward modeling yielded a more consistent set of tapering parameters. We emulated CASA's dirty beam formulation routine\footnote{\url{https://github.com/ryanaloomis/beams_and_weighting}} to compare the fitted \clean\ beam (with FWHMs $\theta_{a}^\prime$ and $\theta_{b}^\prime$) to a \clean\ beam with the desired parameters ($\theta_a$, $\theta_b$, $\phi$). We desired a circularized beam such that $\theta_a = \theta_b = \theta_T$, with sizes $\theta_T = \{0\farcs15, 0\farcs2, 0\farcs3\}$ for Band 6 observations and $\theta_T = \{0\farcs3, 0\farcs5 \}$ for Band 3 observations.

Because density weighting also affects beam shape, there are many combinations of \texttt{robust} values and tapering profiles that deliver the same circularized \clean\ beam profile. To preserve point source sensitivity without introducing large sidelobes, we calculated tapering profiles using a starting value of \texttt{robust=0.5}. We used \texttt{scipy.optimize.minimize} \citep{scipy} to find the best-fitting set of beam tapering parameters that minimized the fit metric 
\begin{equation}
    f(\Theta_a, \Theta_b, \phi_T) = (\theta_t - \theta_{a}^\prime)^2 + (\theta_t - \theta_{b}^\prime)^2.
\end{equation}
If no solution could be found that delivered a beam within 95\% - 100\% of the target resolution, we iterated by reducing the \texttt{robust} value by \texttt{0.25} (towards a uniformly weighted beam) and tried the fitting procedure again. For most Band 6 observations (including those shown in Figure~\ref{fig:tapering}), we were able to successfully find tapered beams using \texttt{robust=0.5} values. For several of the Band 3 observations, however, lower robust values were required to achieve pre-tapered beams with $\theta_a \leq 0\farcs3$. An example of the visibility tapering profiles needed to achieve $\theta_T = \{0\farcs15, 0\farcs3\}$ are shown in the fourth and fifth rows of Figure~\ref{fig:tapering}, respectively. As a final step (not shown in Figure~\ref{fig:tapering}), the images were smoothed the remaining $< 5\%$ to their target resolution using the CASA \texttt{imsmooth} task. By carrying out the bulk of the tapering in the \uv-plane (instead of entirely on the final image), the \clean\ algorithm is able to build a more accurate \clean\ model during the deconvolution process and thus improve final image fidelity.

While our tapering profiles were successful in standardizing beam sizes across the MAPS data products, it is important to realize that even tapered dirty beams whose main lobes deliver circular Gaussian \clean\ beams still have an extended asymmetric shelf. For example, consider the 2D PSF representation of the $0\farcs15$ tapered beam in the fourth row, third column of Figure~\ref{fig:tapering}. This means that any astrophysical flux still remaining in the residual map will retain the asymmetric features of the dirty beam. 

\section{Signal to noise in MAPS products}
\label{sec:signal-to-noise}
The MAPS spectral setup \citep[Table 4]{oberg20} covered more than 40 molecular transitions of interest, many of which are at or near the detection threshold in the five protoplanetary disks we targeted. Given the diverse scientific goals of the MAPS LP, we employed several algorithms to detect and characterize the emission. In this section we discuss the general principles behind the quantification of signal to noise ratio (S/N) and how these apply to several core data products from the large program.

Colloquially, S/N is usually quoted as a one-dimensional quantity in multiples of a $\sigma$ value corresponding to a fractional probability of the Gaussian distribution, e.g., 2$\sigma$ or 95\% (regardless of whether the posterior distribution itself is actually Gaussian). Model assumptions are key to contextualizing any S/N statistic, since it is these (often hidden) assumptions that define the likelihood function, prior distributions, and thus the posterior distribution of the parameter(s) of interest. 

\paragraph{Signal to noise in detection experiments}
Detection experiments are usually discussed in terms of a one-dimensional posterior probability distribution of the amplitude $a$ of the source $p(a\,|\,\mathrm{data})$. The detection S/N is the probability that the inferred amplitude of the source is greater than 0, $p(a > 0\,|\,\mathrm{data})$. In truth, the full posterior distribution is likely multivariate because the model usually has hidden parameters $b$, $c$, etc. In the best situations, the uncertainty pertaining to other model parameters is marginalized out
\begin{equation}
    p(a\,|\,\mathrm{data}) = \int p(a, b, c\,|\, \mathrm{data})\,\mathrm{d}b\, \mathrm{d}c.
\end{equation}
However, for computational or implementation reasons, it may be difficult to explore the probability distributions of these other parameters, and the S/N is estimated using the  \emph{conditional} posterior distribution $p(a\,|\, b, c, \mathrm{data})$ with all other model parameters held fixed. This may overestimate the detection S/N when the model varies significantly under reasonable choices for the other parameters.

Consider the scenario of detecting a point source against a blank background with Gaussian noise. We assume a simple model:  $\delta$-function with known position but unknown amplitude. Assuming the beam is well characterized, the amplitude posterior is defined by a Gaussian centered on the value of flux measured at the location of the point source. The width of the posterior Gaussian corresponds to the thermal RMS of the image. The fraction of the posterior corresponding to fluxes greater than 0 defines the S/N detection probability (e.g., 2$\sigma$ or 95\%). Most realistic scenarios quickly diverge from this idealized scenario to involve more complex models (for example, if the location of the $\delta$-function is not known, then $l$ and $m$ position must be marginalized over and the significance of any particular candidate is diminished).

\subsection{Spatially resolved emission}
When considering spatially resolved emission, it is important to consider the model assumptions that undergird the interpretation of S/N. If the model is misspecified (e.g., a point source when the emission is in fact diffuse) then the S/N calculation, conditional on those assumptions, may not reflect the S/N calculation one actually desires. The more closely a model matches reality, generally speaking, the higher significance a detection that can be achieved. If the model is overly flexible, however, then unknown parameters of the search space (e.g., spatial location, rest frequency) must be marginalized over and the S/N of the detection will suffer. One way to gain intuition for model sensitivity is to attempt to fit the data with a range of different model assumptions and effectively explore some of the hidden parameters, albeit in a limited manner.

\paragraph{The matched filter}
The application of a matched filter to detect spatially resolved but weak line emission from a protoplanetary disk in Keplerian rotation nicely demonstrates some of the model complexity trade-offs \citep[e.g.,][]{loomis18}. In such a framework, a template (such as uniform surface brightness inside some Keplerian pixel mask, e.g., Table~\ref{tab:mask_parameters_13CO}) is assumed, Fourier transformed, and cross-correlated against all available frequency channels. The template amplitude is the sole free parameter of the model. Since the matched filter is linear, the interpretation of the filter response relative to a signal-free region ($\sigma$) is directly equivalent to the point source detection scenario discussed above, and carries many of the same caveats \citep{ruffio17,loomis18}. 

\begin{figure}
    \centering
    \includegraphics[width=\hsize]{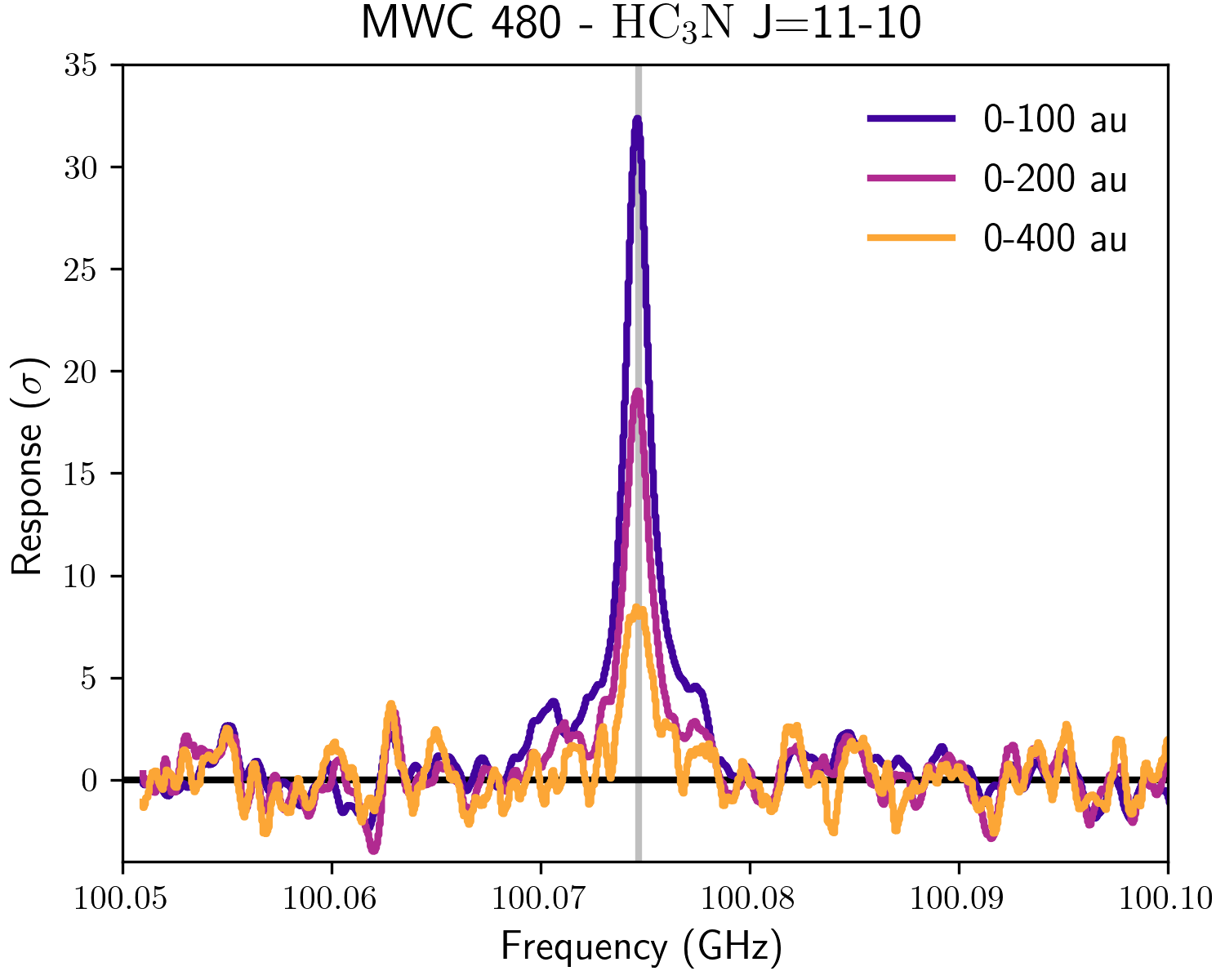}
    \caption{The response of Keplerian matched filter templates with outer radii 100\,au, 200\,au, and 400\,au applied to the MWC~480 \ce{HC_3N} J=11-10 transition measurement set. The detection significance is higher when the template more accurately matches the true spatial distribution of the (inherently compact) emission.}
    \label{fig:filter_masks}
\end{figure}

Detection significance will be maximized when the template is an accurate representation of reality; in most real-world applications an imperfect template will reduce the significance of a detection.  This is demonstrated to some degree in Figure~\ref{fig:filter_masks} with the matched filter applied to the \ce{HC_3N} $J=11-10$ transition in MWC~480, using templates generated from Keplerian masks with different  outer radii (100\,au, 200\,au, and 400\,au). The line is detected with all 3 templates; however, the significance is maximized using the template with the smallest radial extent (100\,au), which best matches the emission morphology of this weak, compact line. Since even the 100\,au template is misspecified to some degree (real disk emission does not appear uniform within some Keplerian mask, but rather radially varies in intensity), the matched filter detections most likely represent lower limits to the maximal S/N detection that could be achieved with an ideal template. More details on the matched filter procedure applied to MAPS data are provided in \citet{ilee20}.

\paragraph{Shifting and stacking}
To aid in the detection and characterization of weak lines, we used the shift and stack technique \citep{yen16,teague16,matra17} on several image cubes. In this process, the Doppler shift corresponding to the Keplerian rotation of the protoplanetary disk is removed from each position-position-velocity pixel in the data cube and each channel in the cube is summed across all pixels. This coherently sums emission across the data cube and results in a higher S/N detection compared to a traditional spatially-integrated spectrum. For a full description of the shift and stack technique applied to MAPS data, as well as the calculation of the uncertainties that result from the deprojection and aggregation of the data, see \citet{ilee20} and \citet{cataldi20}. 

\begin{figure*}
    \centering
    \includegraphics[width=\hsize]{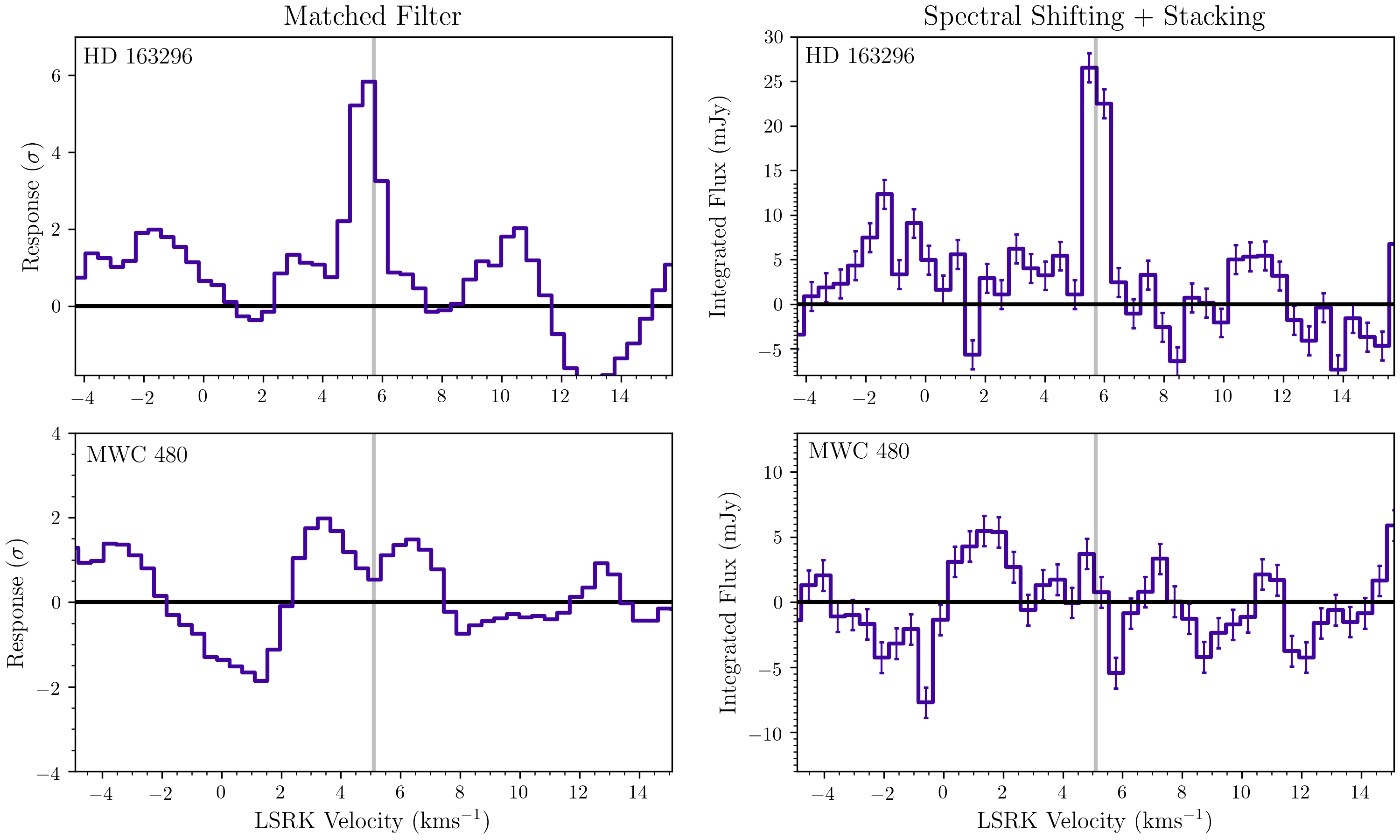}
    \caption{A comparison of the matched filter (left column) and spectral shift and stack (right column) detection techniques applied to the \ce{H^{13}CO^+} $J=1-0$ transition in HD~163296 (top row) and MWC~480 (bottom row). In HD~163296, the transition is detected by both techniques at similar significance (relative to off-source velocities), but in MWC~480 it is not detected by either technique, demonstrating consistency between the two methods.}
    \label{fig:filter_stack}
\end{figure*}

An important benefit to using multiple techniques to detect and characterize faint molecular emission is that confidence builds when independent techniques return consistent results. In Figure~\ref{fig:filter_stack} we compare the results of the matched filter and spectral shift and stack technique applied to the \ce{H^{13}CO^+} $J=1-0$ transition in the HD~163296 and MWC~480 disks \citep[for more information on this and the \ce{HCO^+} $J=1-0$ transitions, see][]{aikawa20}. The \ce{H^{13}CO^+} emission is faint---it cannot be recovered by visual inspection of the channel maps for either disk. Encouragingly, the emission is detected in HD~163296 using both the matched filter (with Keplerian mask template $r_\mathrm{out}=400$\,au radius) and spectral shift and stack technique, while the line is not detected in MWC~480 using either technique. 

The spectral shifting and stacking technique is useful to measure the disk integrated flux and radial intensity profile \citep[especially for transitions with hyperfine structure, e.g.,][]{bergner20,cataldi20,guzman20}, while the matched filter is a more efficient way to search the full MAPS LP for weak transitions (instead of imaging the entire data set). We applied the matched filter to all of the spectral windows to search for any additional lines that were not primary science targets and serendipitously detected satellite lines of \ce{c-C_3H_2} and \ce{C_2H} \citep{ilee20,guzman20}. For a full list of the molecules detected by the MAPS LP, see \citet[Tables 2 \& 3]{oberg20}. 

\section{Summary}
\label{sec:summary}

The MAPS large program represented a significant effort to calibrate and image a large volume of molecular line emission for five protoplanetary disks. In this work, we described the non-Gaussian dirty beam that can arise from multi-configuration ALMA observations, and, after reviewing the \clean ing process, the challenges it presented for accurate flux recovery of faint, extended features. We chose to remedy this issue by implementing the ``JvM correction'' originally proposed by \citet{jorsater95}, which properly scales the \clean\ residual map into consistent units of \jycleanbeam. We also described how we generated custom Keplerian \clean\ masks to guide the deconvolution process, and discussed some imaging artefacts that can result from non-ideal \tclean\ parameters, such as ``stippling.'' To aid in the comparison of different molecular transitions, we also produced image products using beam profiles tapered to common resolutions of FWHM $0\farcs15$, $0\farcs2$, $0\farcs3$, and $0\farcs5$. Finally, we briefly discussed the interpretation of signal to noise and detection significance across the MAPS data products.

We centralized our data processing pipeline on the North American ALMA Science Center (NAASC) computing cluster, located in Charlottesville, VA, USA. Python scripts documenting the reduction and imaging procedures are available.\footnote{\url{http://www.alma-maps.info}} During times of heavy development, we availed ourselves of multiple 16-core machines for days at a time. Including development and incremental reprocessing efforts, we estimate that we utilized 1 year's worth of core hours to produce the MAPS LP image products (i.e., two 16-core machines fully utilized for two weeks). Though still small compared to the computational demands of protoplanetary disk hydrodynamical simulations, for example, this represents a considerably larger computational demand compared to most ALMA observations.

\section{Imaging Data Products} 
\label{sec:data_products}
Here we enumerate the molecular line imaging data products produced from the MAPS LP, which are accessible through a portal to the ALMA archive available after peer review. For a detailed description of the naming conventions for all MAPS LP image products, see \citet[\S3.5,][]{oberg20}. For a description of the continuum-only MAPS image products, see \citet{sierra20}.

For each combination of disk \citep[e.g., MWC~480; see Table 1,][]{oberg20} and transition \citep[e.g., DCN $J=3-2$; see Tables 2 \& 3,][]{oberg20}, the archive contains two minimal measurement sets produced with the \texttt{cvel2} and \texttt{split} tasks with visibilities pertaining to that disk and transition pair. The first contains the visibilities including continuum emission, while the second contains the line visibilities with the continuum subtracted. During the invocation of \texttt{cvel2}, we coarsened the spectral channels slightly from their native spacings (see Table 4 of \citet{oberg20}) to a uniform set of channels spaced 0.5 km~s$^{-1}$ apart in B3 and 0.2 km~s$^{-1}$ in B6.

The archive also contains a set of image products generated from each measurement set using various beams. For the Band 3 transitions, these beams are untapered \texttt{robust=0.5}, tapered 0\farcs30, and tapered 0\farcs50. For the Band 6 transitions, these beams are untapered \texttt{robust=0.5}, tapered 0\farcs15, tapered 0\farcs20, and tapered 0\farcs30. For a full description of beam sizes available in each band, see Table 5, \citet{oberg20}. For each beam setting, the following image products are available:
\begin{itemize}
\item Keplerian binary \clean\ mask cube
\item (unconvolved) \clean\ model cube
\item residual cube from \clean ing process (before JvM correction)
\item \clean ed image cube under standard workflow (Figure~\ref{fig:workflow}), with and without primary beam correction
\item \clean ed image cube under JvM correction (Figure~\ref{fig:workflow}), with and without primary beam correction. The primary beam and ``JvM''--corrected cube is the recommended data cube for most scientific use cases.
\item A Python script to reproduce the data products from the minimal continuum-subtracted measurement set.
\end{itemize}

The following is a typical workflow to produce a set of image products. We used the \tclean\ task with multiscale \clean\ and \texttt{scales=[0, 5, 15, 25]} pixels,  where  the  pixel  size  was  chosen  to  correspond to $\approx 1/7$th of the beam FWHM. For all lines except $\mathrm{CO}$, we used Keplerian \clean\ masks matched to the ${}^{13}\mathrm{CO}$ $J=2-1$  emission. We iterated the \clean\ algorithm such that the peak residual emission was below a threshold of $4\times\mathrm{RMS}$. For untapered beams we used Briggs weighting of \texttt{robust=0.5}. For tapered beams we forward modeled the CASA beam fitting process to calculate the value of the \texttt{uvtaper} argument that achieves the target resolution using the largest (most natural) \texttt{robust} value still $\leq 0.5$. Finally, we calculated the JvM factor $\epsilon$ via the ratio of the \clean\ beam volume to the dirty beam volume, scaled the residual map by $\epsilon$ and summed it with the convolved \clean\ model to produce the JvM-corrected image cube. We recommend consulting the Python script accompanying each set of image products for the specific \tclean\ parameters used to generate a particular image product.

For more information on the additional value added data products (VADP) like moment maps, radial profiles, and emission surfaces provided for most transitions across a range of beam sizes see \citet{law20_rad, law20_surf}.

\acknowledgments
This paper makes use of the following ALMA data: ADS/JAO.ALMA\#2018.1.01055.L. ALMA is a partnership of ESO (representing its member states), NSF (USA) and NINS (Japan), together with NRC (Canada), MOST and ASIAA (Taiwan), and KASI (Republic of Korea), in cooperation with the Republic of Chile. The Joint ALMA Observatory is operated by ESO, AUI/NRAO and NAOJ. The National Radio Astronomy Observatory is a facility of the National Science Foundation operated under cooperative agreement by Associated Universities, Inc. We thank the NAASC for the use of computational resources on the North American ALMA Science Center (NAASC) computing cluster.

C.J.L. acknowledges funding from the National Science Foundation Graduate Research Fellowship under Grant DGE1745303. R.T. acknowledges support from the Smithsonian Institution as a Submillimeter Array (SMA) Fellow. K.I.\"O. acknowledges support from the Simons Foundation (SCOL \#321183) and an NSF AAG Grant (\#1907653). J.H. and S.M.A. acknowledge funding support from the National Aeronautics and Space Administration under Grant No. 17-XRP17 2-0012 issued through the Exoplanets Research Program. J. H. acknowledges support for this work provided by NASA through the NASA Hubble Fellowship grant \#HST-HF2-51460.001-A awarded by the Space Telescope Science Institute, which is operated by the Association of Universities for Research in Astronomy, Inc., for NASA, under contract NAS5-26555. I.C. was supported by NASA through the NASA Hubble Fellowship grant HST-HF2-51405.001-A awarded by the Space Telescope Science Institute, which is operated by the Association of Universities for Research in Astronomy, Inc., for NASA, under contract NAS5-26555. Y.A. acknowledges support by NAOJ ALMA Scientific Research Grant Code 2019-13B, and Grant-in-Aid for Scientific Research 18H05222 and 20H05847. R.L.G. acknowledges support from a CNES fellowship grant. K.Z. acknowledges the support of the Office of the Vice Chancellor for Research and Graduate Education at the University of Wisconsin – Madison with funding from the Wisconsin Alumni Research Foundation, and support of the support of NASA through Hubble Fellowship grant HST-HF2-51401.001. awarded by the Space Telescope Science Institute, which is operated by the Association of Universities for Research in Astronomy, Inc., for NASA, under contract NAS5-26555. C.W. acknowledges financial support from the University of Leeds, STFC and UKRI (grant numbers ST/R000549/1, ST/T000287/1, MR/T040726/1). J.D.I. acknowledges support from the Science and Technology Facilities Council of the United Kingdom (STFC) under ST/T000287/1. K.R.S. acknowledges the support of NASA through Hubble Fellowship Program grant HST-HF2-51419.001, awarded by the Space Telescope Science Institute,which is operated by the Association of Universities for Research in Astronomy, Inc., for NASA, under contract NAS5-26555. G.C. is supported by NAOJ ALMA Scientific Research Grant Code 2019-13B. J.B.B. acknowledges support from NASA through the NASA Hubble Fellowship grant \#HST-HF2-51429.001-A, awarded by the Space Telescope Science Institute, which is operated by the Association of Universities for Research in Astronomy, Inc., for NASA, under contract NAS5-26555. J.B. acknowledges support by NASA through the NASA Hubble Fellowship grant \#HST-HF2-51427.001-A awarded  by  the  Space  Telescope  Science  Institute,  which  is  operated  by  the  Association  of  Universities  for  Research  in  Astronomy, Incorporated, under NASA contract NAS5-26555. Y.Y. is supported by IGPEES, WINGS Program, the University of Tokyo. H.N. acknowledges support by NAOJ ALMA Scientific Research Grant Code 2018-10B and Grant-in-Aid for Scientific Research 18H05441. T.T. is supported by JSPS KAKENHI Grant Numbers JP17K14244 and JP20K04017. FMe acknowledges support from ANR of France under contract ANR-16-CE31-0013 (Planet-Forming-Disks)  and ANR-15-IDEX-02 (through CDP ``Origins of Life''). L.M.P.\ acknowledges support from ANID project Basal AFB-170002 and from ANID FONDECYT Iniciaci\'on project \#11181068. A.S.B acknowledges the studentship funded by the Science and Technology Facilities Council of the United Kingdom (STFC). A.D.B. acknowledges support from NSF AAG Grant \#1907653. E.A.B. acknowledges support from NSF AAG Grant \#1907653. L.I.C. gratefully acknowledges support from the David and Lucille Packard Foundation and Johnson \& Johnson's WiSTEM2D Program. V.V.G. acknowledges support from FONDECYT Iniciaci\'on 11180904 and ANID project Basal AFB-170002. A.R.W. acknowledges support from the Virginia Space Grant Consortium and the National Science Foundation Graduate Research Fellowship Program under Grant No. DGE1842490. N.T.K. acknowledges support provided by the Alexander von Humboldt Foundation in the framework of the Sofja Kovalevskaja Award endowed by the Federal Ministry of Education and Research.

\vspace{5mm}
\facilities{ALMA}

\software{astropy \citep{astropy1,astropy2}, CASA \citep[v6.1.0;][]{mcmullin07}, GoFish \citep{teague19}, NumPy \citep{numpy}, SciPy \citep{scipy}, \texttt{keplerian\_mask} \citep{teague_kepmask}}

\bibliography{MAPS}{}
\bibliographystyle{aasjournal}

\end{document}